\documentclass[twocolumn,english,amsart,showpacs,preprintnumbers,amsmath,amssymb,floatfix]{revtex4-1}
\usepackage{tikz,xcolor}
\usepackage[colorlinks = true,
linkcolor = blue,
urlcolor  = blue,
citecolor = blue,
anchorcolor = blue]{hyperref}

\definecolor{lime}{HTML}{A6CE39}
\DeclareRobustCommand{\orcidicon}{%
	\begin{tikzpicture}
		\draw[lime, fill=lime] (0,0)
		circle [radius=0.16]
		node[white] {{\fontfamily{qag}\selectfont \tiny ID}};
		\draw[white, fill=white] (-0.0625,0.095)
		circle [radius=0.007];
	\end{tikzpicture}
	\hspace{-2mm}
}

\foreach \x in {A, ..., Z}{%
	\expandafter\xdef\csname orcid\x\endcsname{\noexpand\href{https://orcid.org/\csname orcidauthor\x\endcsname}{\noexpand\orcidicon}}
}

\usepackage[T1]{fontenc}
\usepackage[latin9]{inputenc}
\usepackage{color}
\usepackage{array}
\usepackage{amstext}
\usepackage{graphicx}
\usepackage{esint}
\usepackage{rotating}
\usepackage{appendix}
\usepackage{float}

\usepackage[font=small,labelfont=bf]{caption}
\usepackage{xcolor}
\usepackage{ulem}

\makeatletter

\@ifundefined{textcolor}{}
{%
	\definecolor{BLACK}{gray}{0}
	\definecolor{WHITE}{gray}{1}
	\definecolor{RED}{rgb}{1,0,0}
	\definecolor{GREEN}{rgb}{0,1,0}
	\definecolor{BLUE}{rgb}{0,0,1}
	\definecolor{CYAN}{cmyk}{1,0,0,0}
	\definecolor{MAGENTA}{cmyk}{0,1,0,0}
	\definecolor{YELLOW}{cmyk}{0,0,1,0}
}

\@ifundefined{definecolor}
{\usepackage{color}}{}
\@ifundefined{definecolor} 
{\usepackage{color}}{}
\makeatother
\usepackage{babel}

\allowdisplaybreaks

\begin{document}
	

	\title{Form factors of nucleon by using t-dependence
		of parton distribution functions}

	\author{Hossein Vaziri\orcidA{}}
	\email{Hossein.Vaziri@shahroodut.ac.ir}
	
	\author{Mohammad Reza Shojaei\orcidB{}}
	\thanks{Corresponding author}
	\email{Shojaei.ph@gmail.com}

	\affiliation {
		Department of Physics, Shahrood University of Technology, P. O. Box 36155-316, Shahrood, Iran}

	\date{\today}

	%
	%

	%
	\begin{abstract}\label{abstract}
		This paper calculates the elastic form factors for nucleons based on generalized parton distributions using an extended new ansatz introduced in \href{http://dx.doi.org/10.1103/PhysRevC.105.025202}{{\rm Phys. Rev. C} {\bfseries 105}, no.2, 025202 (2022)}. Different parton distribution functions (PDFs) are considered, and modifications are made to the free parameters of the new ansatz. The obtained results are systematically compared among the combinations of different PDFs and ansatzes for high ranges of momentum transfer, with $-t < 35 GeV^{-2}$. The minimum suitable parameters are used to parametrize the model. After obtaining the form factors, we proceed to compute the electric radius and the transversely unpolarized densities for the nucleon. In addition, we derive the parton distribution functions (PDFs) that depend on the impact parameter. Finally, we analyze the results by comparing them with the findings from other research and experimental data.

	\end{abstract}
	%

	
	\maketitle

	%

	\section{Introduction}\label{sec:sec1}

Nucleons can be studied through measuring their elastic form factors, which can be obtained using unique techniques such as proton and neutron elastic scattering tests based on Rosenbluth methods~\cite{Perdrisat:2006hj,Arnold:1980zj,JeffersonLabHallA:2011yyi,Gramolin:2021gln,Punjabi:2005wq,Rosenbluth:1950yq,Gayou:2001qt,Halzen}. Hadron's quark distributions are revealed through GPDs. GPDs and the Dirac and Pauli form factors are connected and are used to study the structure of nucleons. Generally, GPDs depend on the momentum transfer $t=Q^2$, the average longitudinal momentum fraction $x$ of the partons in the hard-scattering, and the skewness parameter $\xi$ which measures the longitudinal momentum transfer~\cite{Radyushkin:2011dh,Guidal:2013rya, Bhattacharya:2019cme,HajiHosseiniMojeni:2022okc}. The electromagnetic form factors in GPDs can be calculated using a variety of techniques and parametrizations~\cite{Radyushkin:1998rt,Goeke:2001tz,Stoler:2001xa,Guidal:2004nd,Mondal:2015uha,Nikkhoo:2015jzi,SattaryNikkhoo:2018gzm,HajiHosseiniMojeni:2022tzn,Selyugin:2009ic,Selyugin:2014sca,Sharma:2016cnf,HajiHosseiniMojeni:2022tzn,Hou:2019efy}. Several investigations have been made into the dependence of GPDs on $x$ and $t$, with suggested ansatzes for the $t$ dependence, while the $x$ dependence is considered to be used for parton distribution functions.
This section is based on selecting an appropriate parton distribution function (PDF) and an ansatz that considers the $x$ and $t$ dependencies of GPDs. By modifying the ansatz parameters and combining the ansatz with the appropriate PDFs, the calculation results of form factors have a suitable agreement with experimental data. We have used the MRST2002~\cite{Martin:2002dr}, JR09~\cite{Jimenez-Delgado:2008orh}, and CT18~\cite{Hou:2019efy} parton distribution functions and applied the modified Gaussian ansatz~\cite{Selyugin:2009ic}, the extended ER~\cite{Guidal:2004nd}, and HS22 ansatz~\cite{HajiHosseiniMojeni:2022tzn} for high momentum transfer ranges to systematically compare the electromagnetic form factors generated from all the analyzed ansatzes using various PDFs.
	Below is a structure for this essay:\\ In Sec.\ref{sec:sec2}, a review of generalized parton distribution theoretical features, elastic form factors, and electric mean squared for the nucleon is studied.\\ 
	In Sec.\ref{sec:sec3}, the theoretical framework of transverse charge and magnetic densities is reviewed.\\ The GPDs in the space of impact parameters are covered in Sec.
	\ref{sec:sec4}.\\Finally, our conclusions are presented in Sec.~\ref{sec:conclusion}.
	\section{GENERALIZED PARTON DISTRIBUTIONS AND
		ELASTIC FORM FACTORS}\label{sec:sec2}
Generalized parton distributions (GPDs) are considered one of the essential techniques for studying the structure of nucleons~\cite{Muller:1994ses,Ji:1996ek,Radyushkin:1997ki,Ji:1996nm,Collins:1996fb}. In order to calculate the spatial distribution of partons in the transverse plane and create a three-dimensional representation of the nucleon, it is necessary to perform deep virtual Compton scattering experiments and apply a Fourier transform to the t-dependence of GPDs~\cite{Burkardt:2002hr,Burkardt:2000za,Ralston:2001xs,Belitsky:2003nz,SattaryNikkhoo:2018odd}.
	
	The elastic form factors, which are also known as the Dirac and Pauli form factors, can be described in terms of GPDs for valence quarks $H_q(x,\xi,t)$ and $E_q(x,\xi,t)$ by using the quark sum rules of the $u$ and $d$ flavors~\cite{Guidal:2004nd,Nikkhoo:2015jzi,RezaShojaei:2016oox}.

	\begin{equation}
		F_1(t) = \sum_q e_q \int_{-1}^{1} dxH_q(x,\xi,t),
		\label{eq:F1xi}
	\end{equation}
	\begin{equation}
		F_2(t) =\sum_q e_q \int_{-1}^{1} dxE_q(x,\xi,t),
		\label{eq:F2xi}
	\end{equation}

When the momentum is transverse and located in the space-like region, the value of $\xi$ is equal to zero. In the range of, $0 < x < 1$, the integration region can be reduced. By revising the elastic form factors, we can obtain~\cite{Guidal:2004nd,Selyugin:2009ic}:
	\begin{equation}
		F_{1}(t)=\sum_{q}e_{q}\int_{0}^{1}dx\mathcal{H}^{q}(x,t),
		\label{eq:F1}
	\end{equation}
	
	\begin{equation}
		F_{2}(t)=\sum_{q}e_{q}\int_{0}^{1}dx {\varepsilon }^{q}(x,t),
		\label{eq:F2}
	\end{equation}

 In any model, there are differences between the functions $\mathcal{H}(x)$ and $\varepsilon(x)$. The $x \rightarrow 1$ limit of $\varepsilon(x)$ should include more powers of $(1 - x)$ than $\mathcal{H}(x)$ to produce a quicker reduction with $t$~\cite{Guidal:2004nd,Selyugin:2009ic}. Therefore, it can be concluded that
	
	\begin{figure*}
		\includegraphics[clip,width=0.45\textwidth]{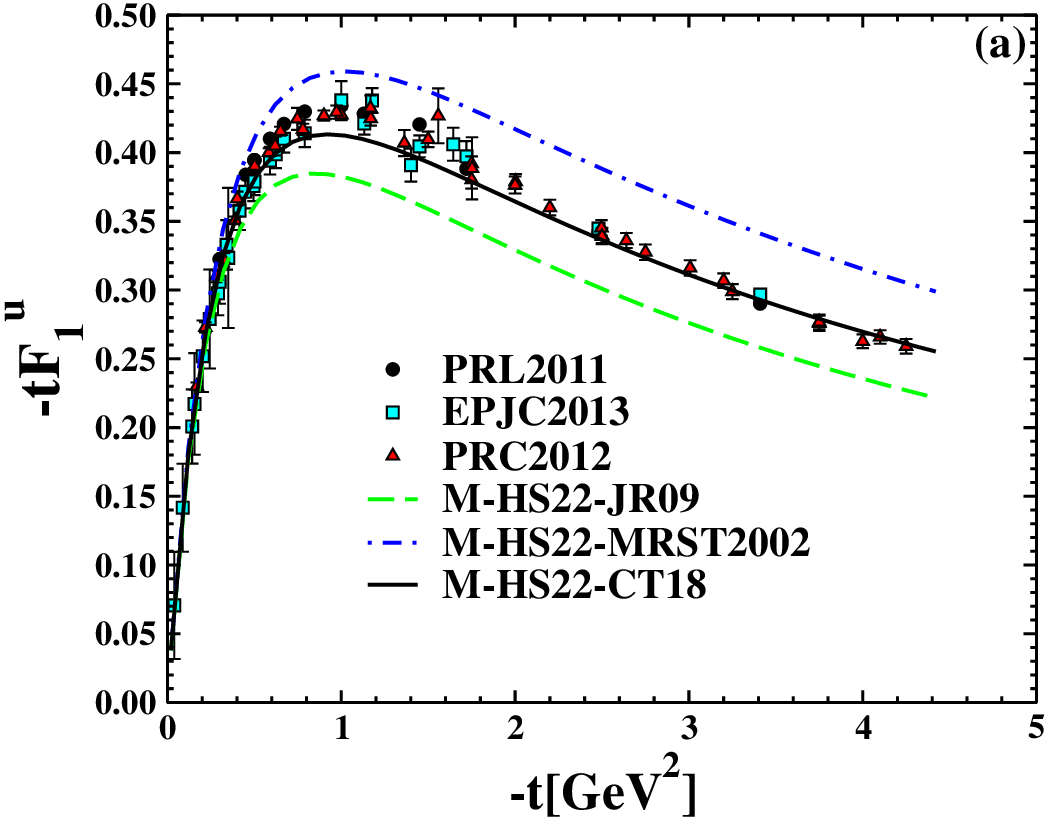}
							\hspace*{2mm}
		\includegraphics[clip,width=0.45\textwidth]{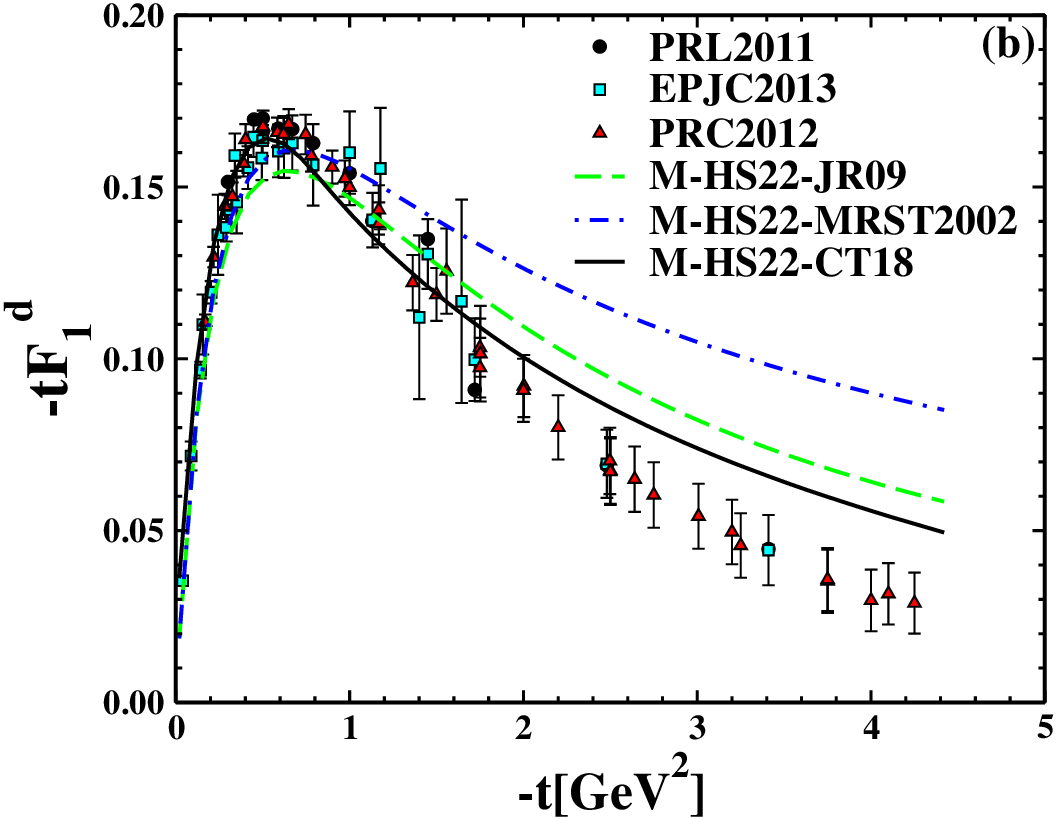}
									\vspace*{3mm}\\
		\includegraphics[clip,width=0.45\textwidth]{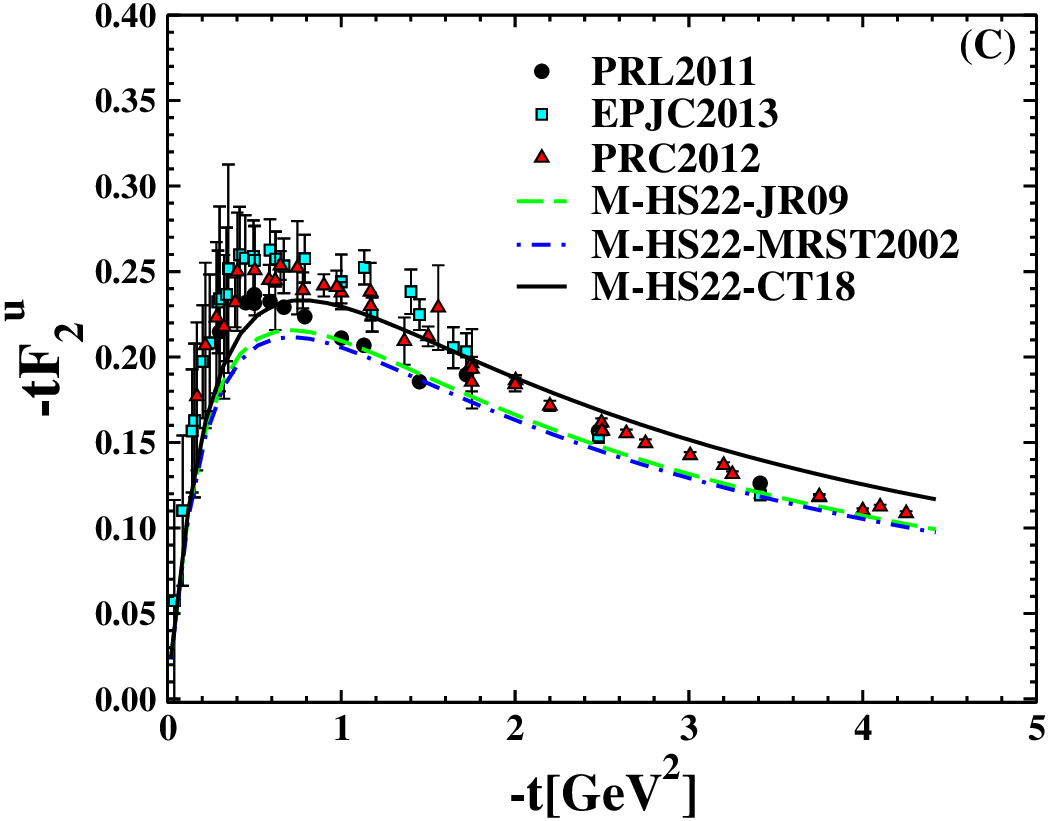}
									\hspace*{2mm}
		\includegraphics[clip,width=0.45\textwidth]{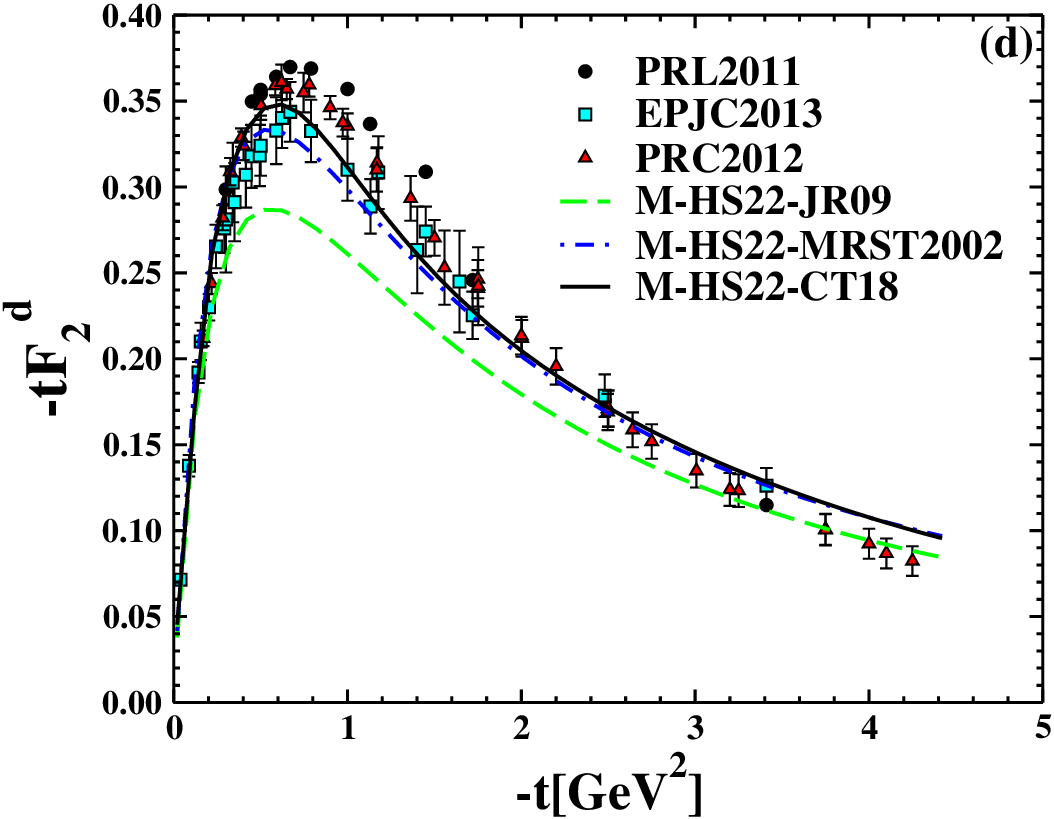}
			\vspace*{1.5mm}

		\caption{\footnotesize  The form factors of $u$ and $d$ quarks multiplied by $t$ as a function of $-t$ using the M-HS22 ansatz and CT18 PDF~\cite{Hou:2019efy},JR09 PDF ~\cite{Jimenez-Delgado:2008orh}, MRST2002 PDF ~\cite{Martin:2002dr}. The points shown are extractions based on experimental data from ~\cite{Qattan:2012zf} (triangle up),~\cite{Cates:2011pz} (circle) and ~\cite{Diehl:2013xca} (square).}
		\label{fig:tfud1}
	\end{figure*}
	
	\begin{figure*}
		\includegraphics[clip,width=0.45\textwidth]{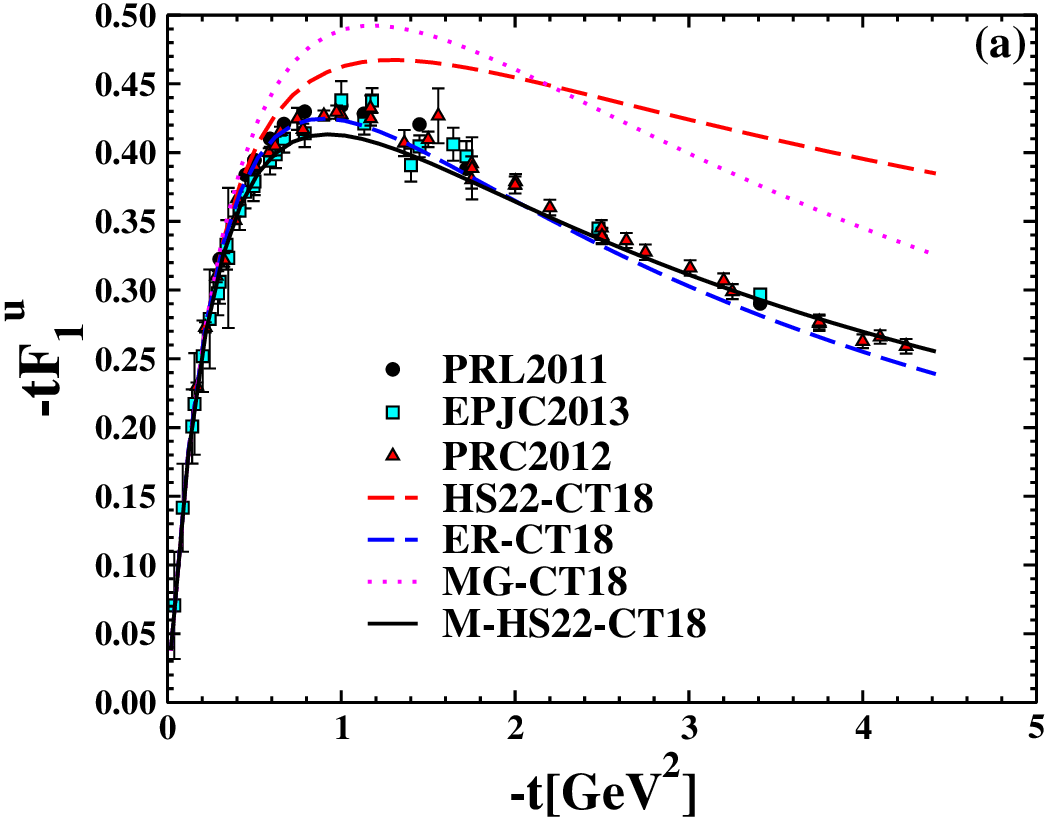}
									\hspace*{2mm}
		\includegraphics[clip,width=0.45\textwidth]{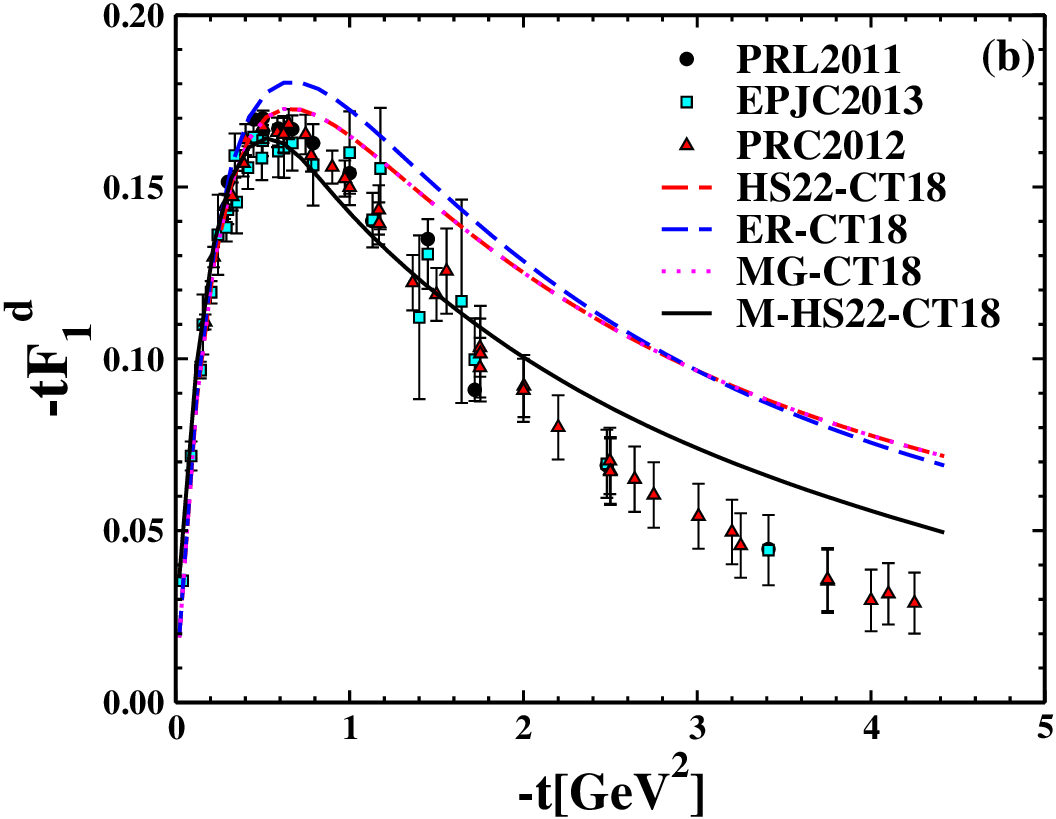}
											\vspace*{3mm}\\
		\includegraphics[clip,width=0.45\textwidth]{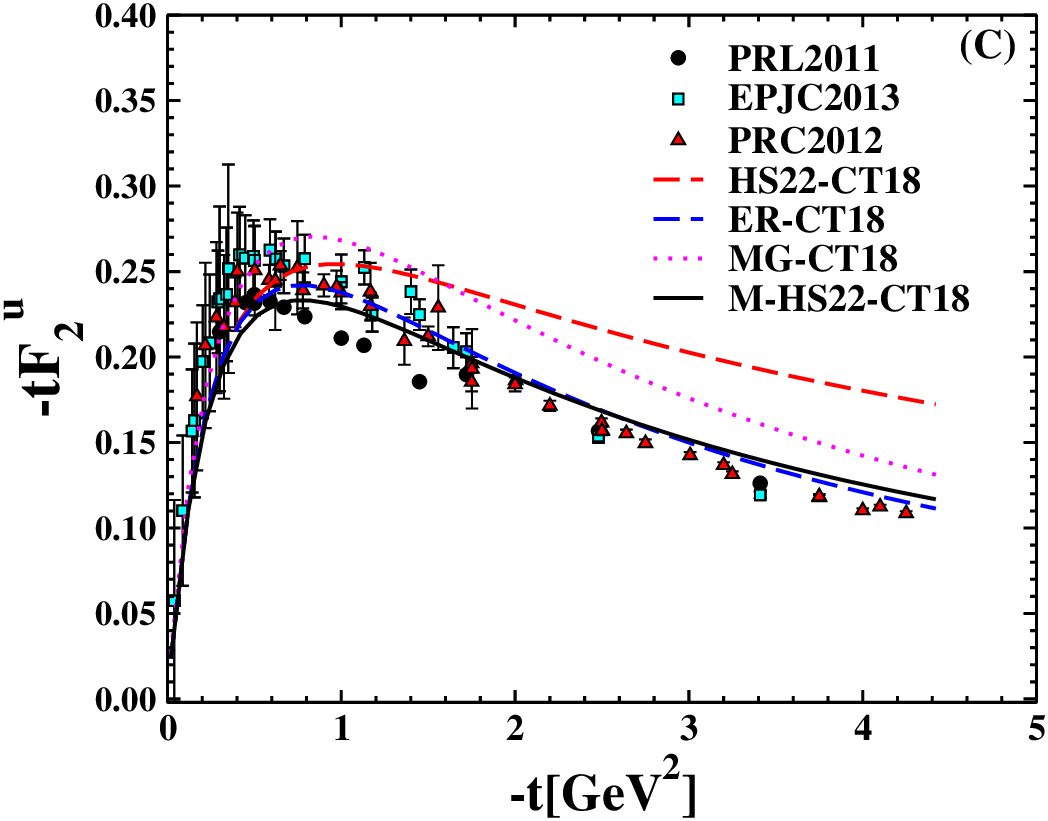}
									\hspace*{2mm}
		\includegraphics[clip,width=0.45\textwidth]{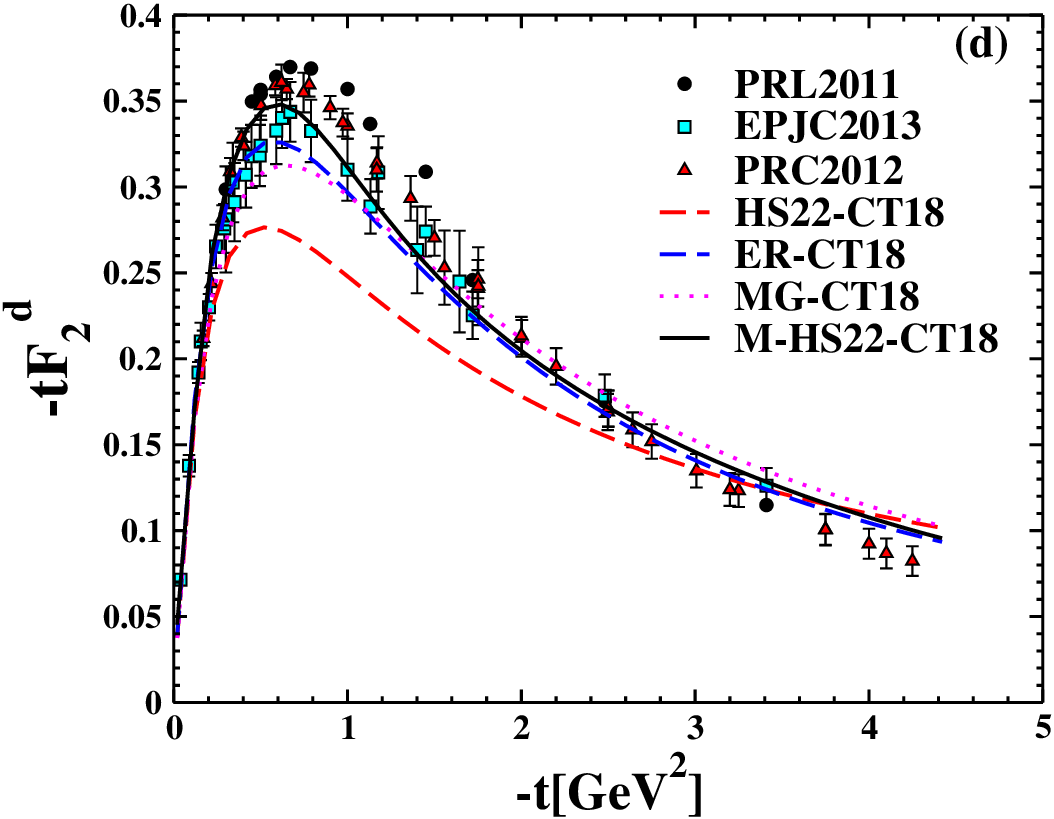}
									\vspace*{1.5mm}
		\caption{\footnotesize  The form factors of $u$ and $d$ quarks multiplied by $t$ as a function of $-t$ using the CT18 pdf ~\cite{Hou:2019efy} and HS22 ansatz~\cite{HajiHosseiniMojeni:2022tzn}, ER ansatz~\cite{Guidal:2004nd}, MG ansatz~\cite{Selyugin:2009ic} and M-HS22 ansatz. The points shown are extractions based on experimental data from ~\cite{Qattan:2012zf} (triangle up),~\cite{Cates:2011pz} (circle) and ~\cite{Diehl:2013xca} (square).}
		\label{fig:tfud}
	\end{figure*}

	\begin{figure*}
		\includegraphics[clip,width=0.45\textwidth]{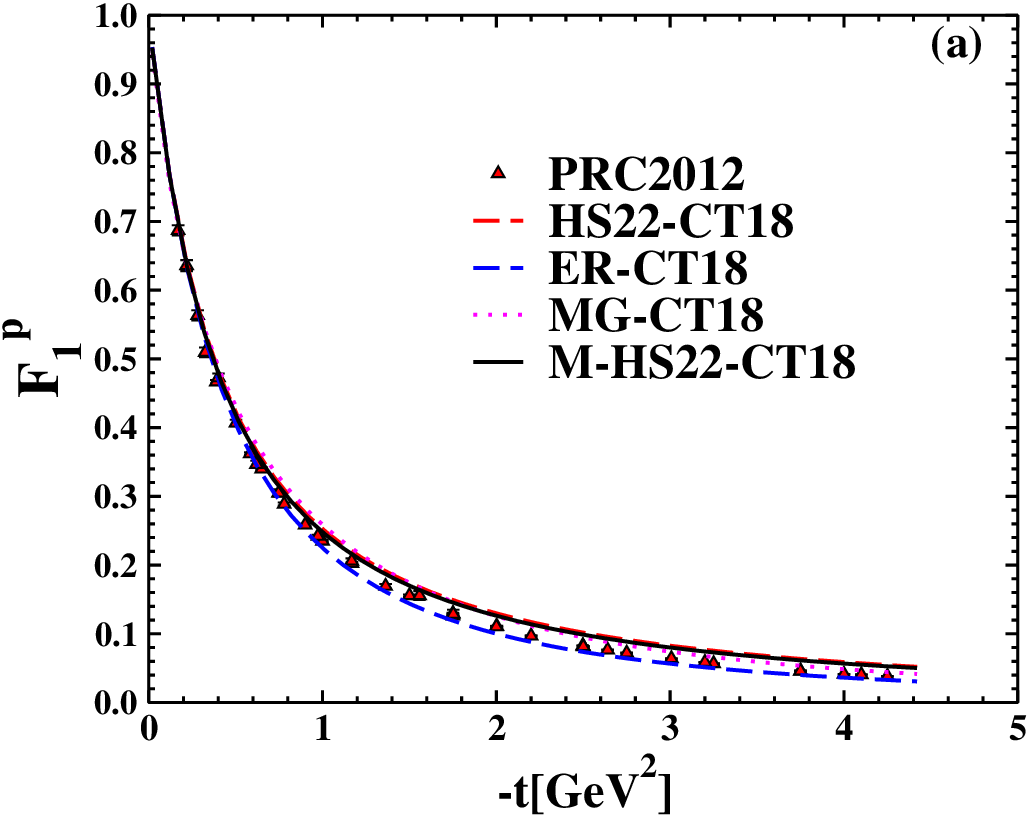}
											\hspace*{2mm}
		\includegraphics[clip,width=0.45\textwidth]{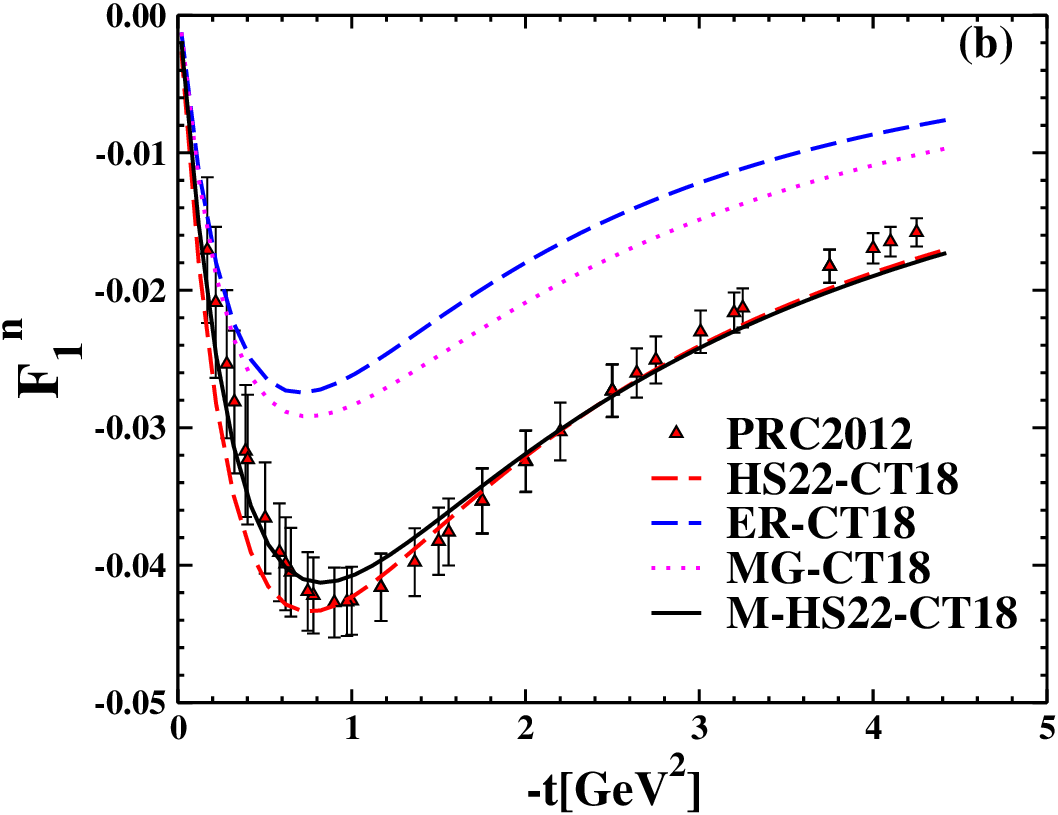}
											\vspace*{3mm}\\
		\includegraphics[clip,width=0.45\textwidth]{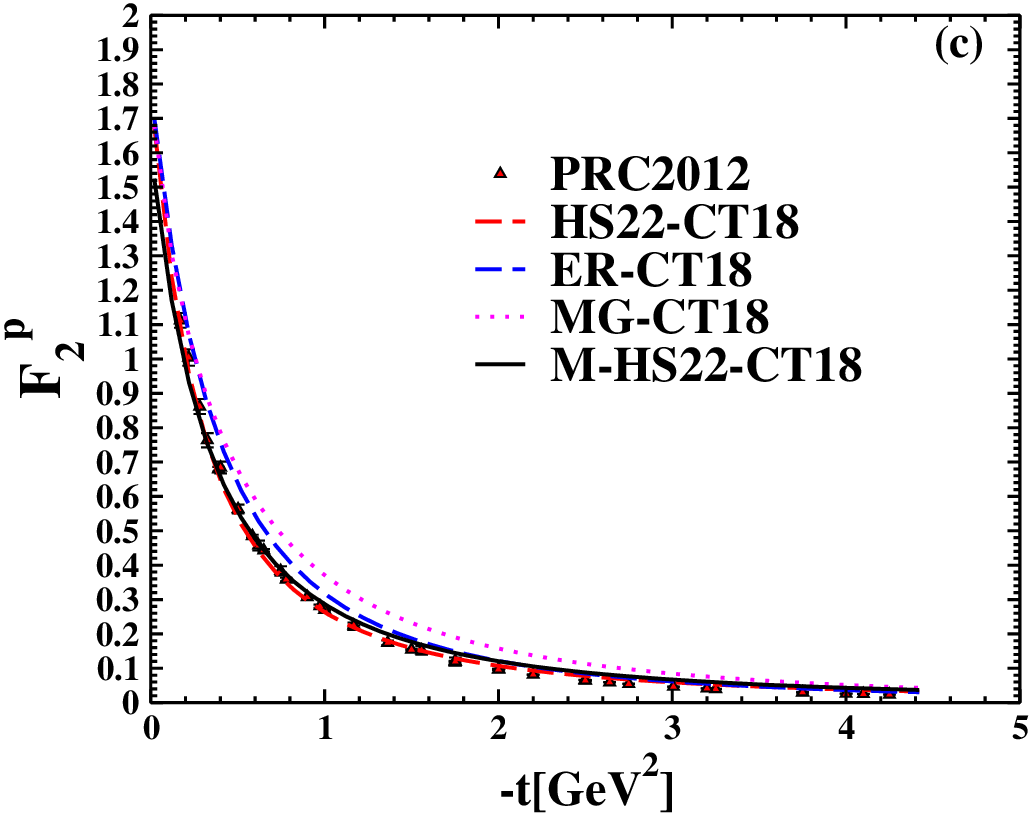}
											\hspace*{2mm}
		\includegraphics[clip,width=0.45\textwidth]{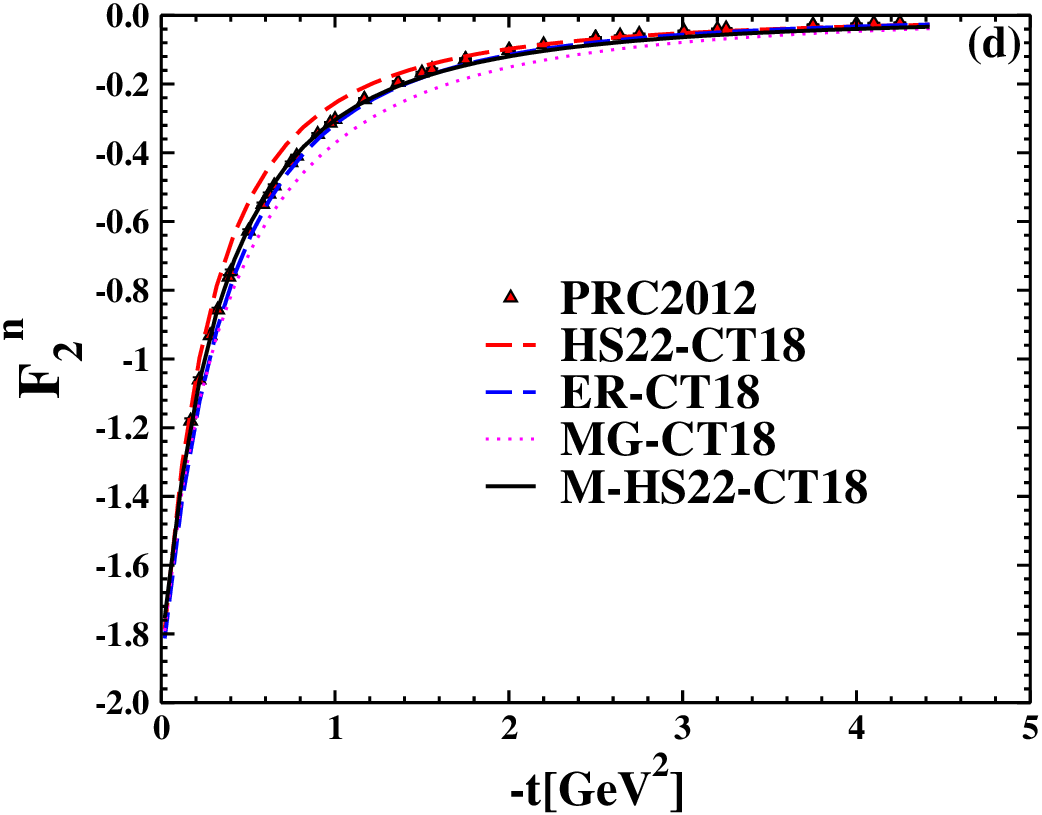}
											\hspace*{1.5mm}

		\caption{\footnotesize    The $F_1^{p,n}$ and $F_2^{p,n}$ as a function of $-t$ using the CT18 pdf ~\cite{Hou:2019efy} and HS22 ansatz~\cite{HajiHosseiniMojeni:2022tzn}, ER ansatz~\cite{Guidal:2004nd}, MG ansatz~\cite{Selyugin:2009ic} and M-HS22 ansatz. The points shown are extractions based on experimental data from ~\cite{Qattan:2012zf} (triangle up).}
		\label{fig:fpn}
	\end{figure*}

	\begin{figure*}
		\includegraphics[clip,width=0.45\textwidth]{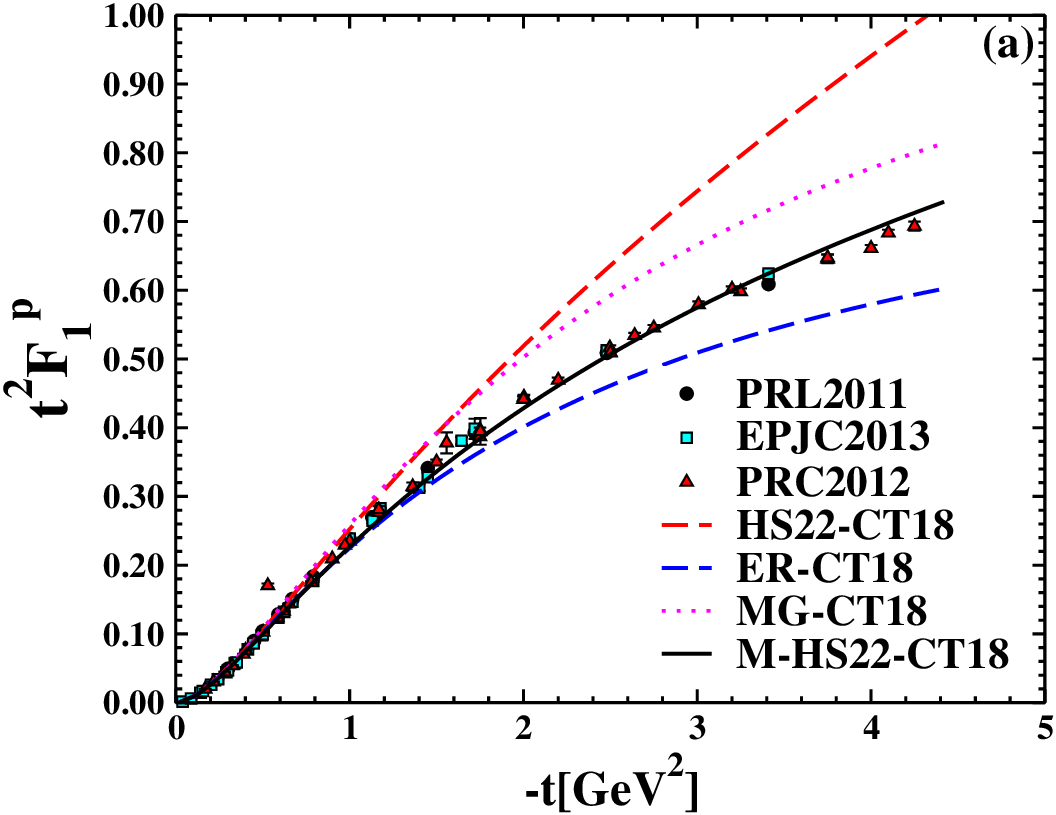}
											\hspace*{2mm}
		\includegraphics[clip,width=0.45\textwidth]{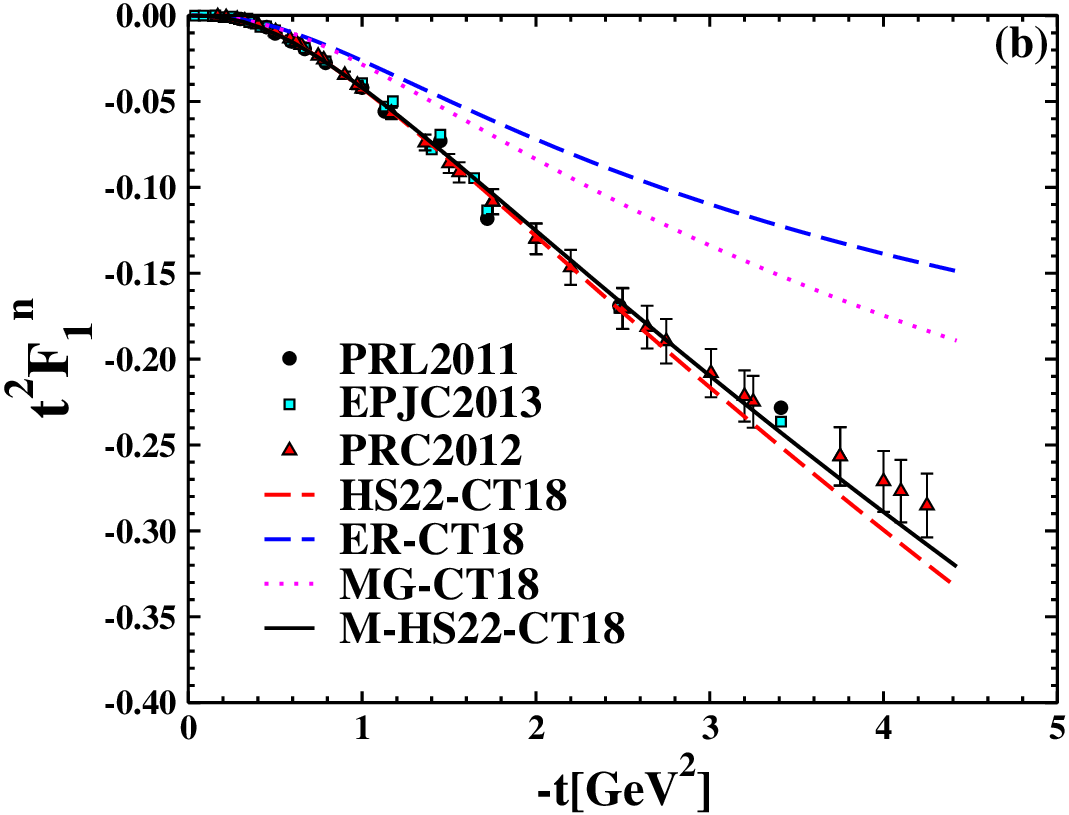}
											\vspace*{3mm}\\
		\includegraphics[clip,width=0.45\textwidth]{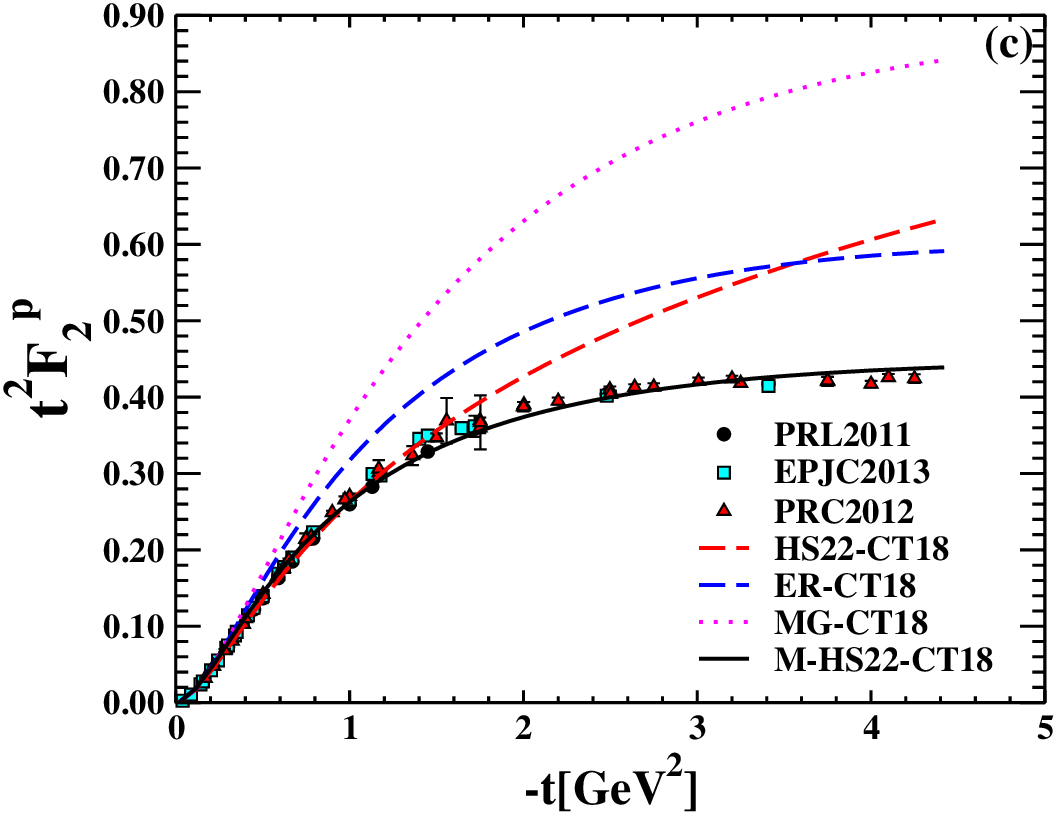}
													\hspace*{2mm}
		\includegraphics[clip,width=0.45\textwidth]{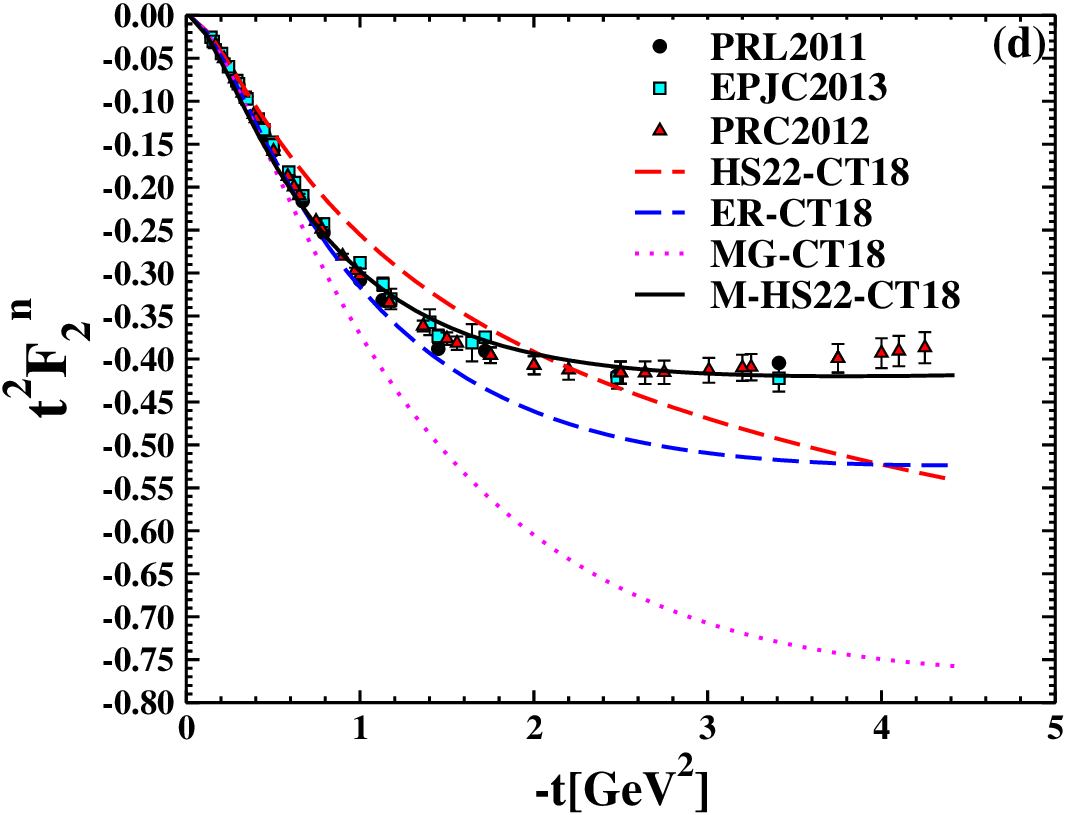}
			\vspace*{1.5mm}
		\caption{\footnotesize  The $t^2F_1^{p,n}$ and $t^2F_2^{p,n}$ form factors as a function of $-t$ using the CT18 pdf ~\cite{Hou:2019efy} and HS22 ansatz~\cite{HajiHosseiniMojeni:2022tzn}, ER ansatz~\cite{Guidal:2004nd}, MG ansatz~\cite{Selyugin:2009ic} and M-HS22 ansatz. The points shown are extractions based on experimental data from ~\cite{Qattan:2012zf} (triangle up),~\cite{Cates:2011pz} (circle) and ~\cite{Diehl:2013xca} (square).}
		\label{fig:t2fud}
	\end{figure*}

	\begin{figure*}
		\includegraphics[clip,width=0.45\textwidth]{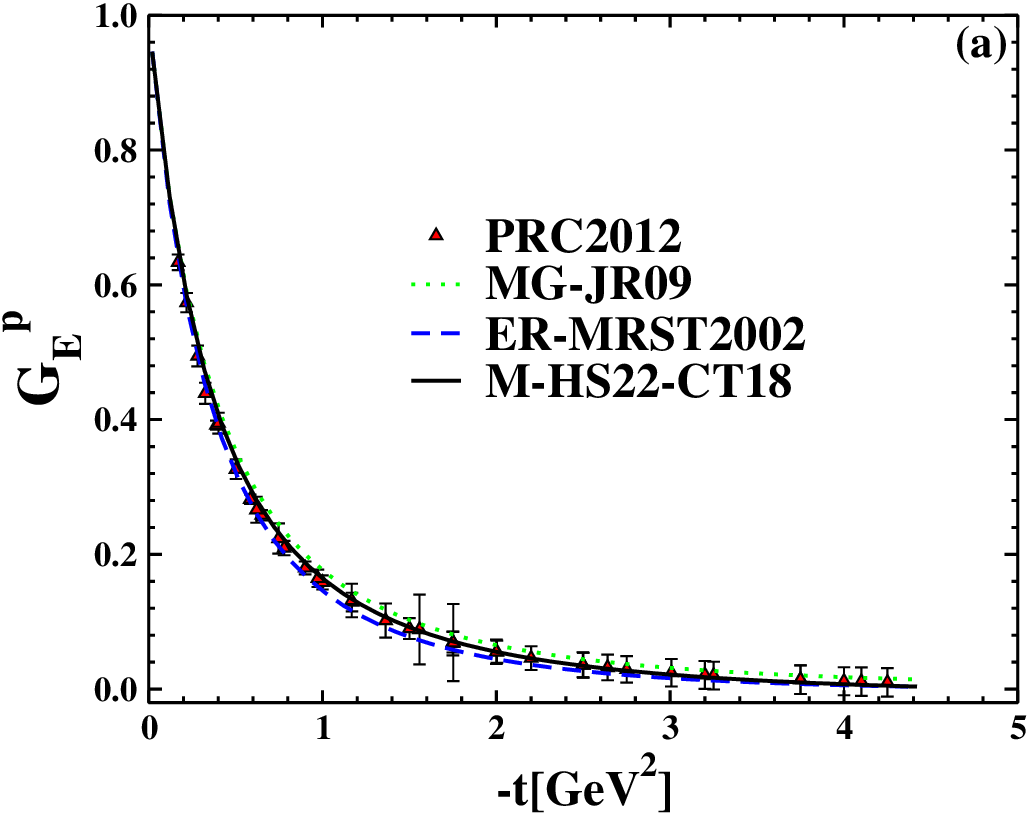}
													\hspace*{2mm}
		\includegraphics[clip,width=0.45\textwidth]{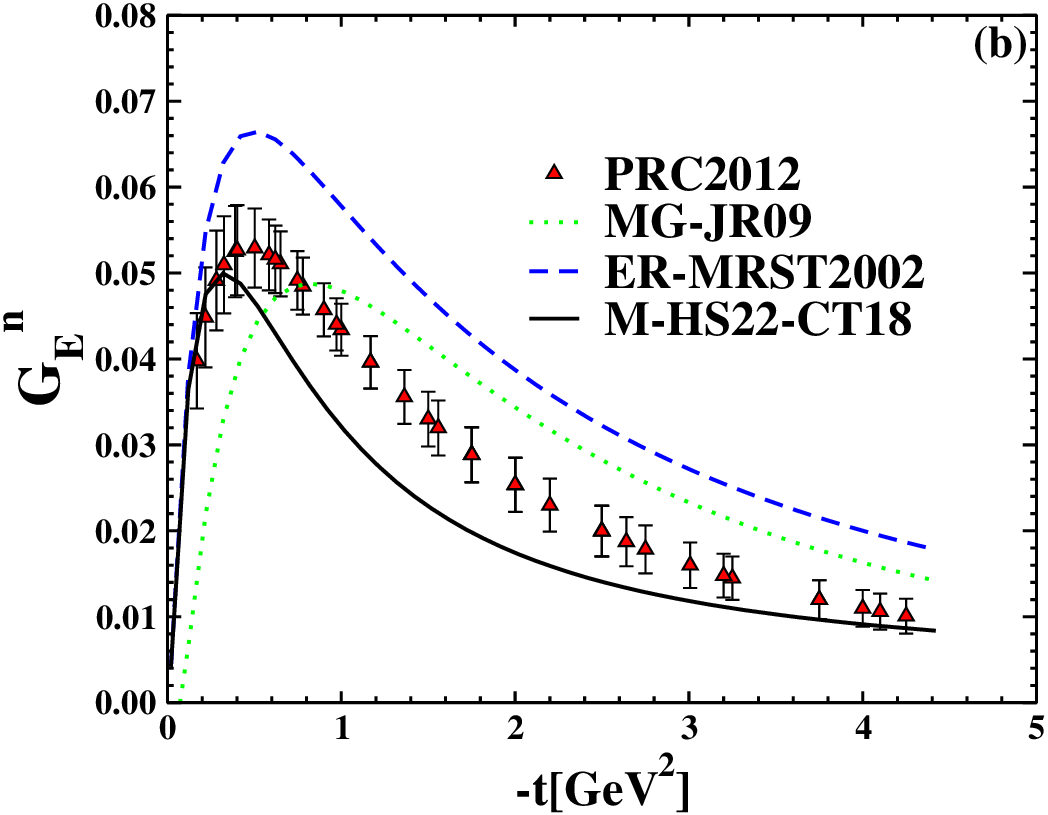}
											\vspace*{3mm}\\
		\includegraphics[clip,width=0.45\textwidth]{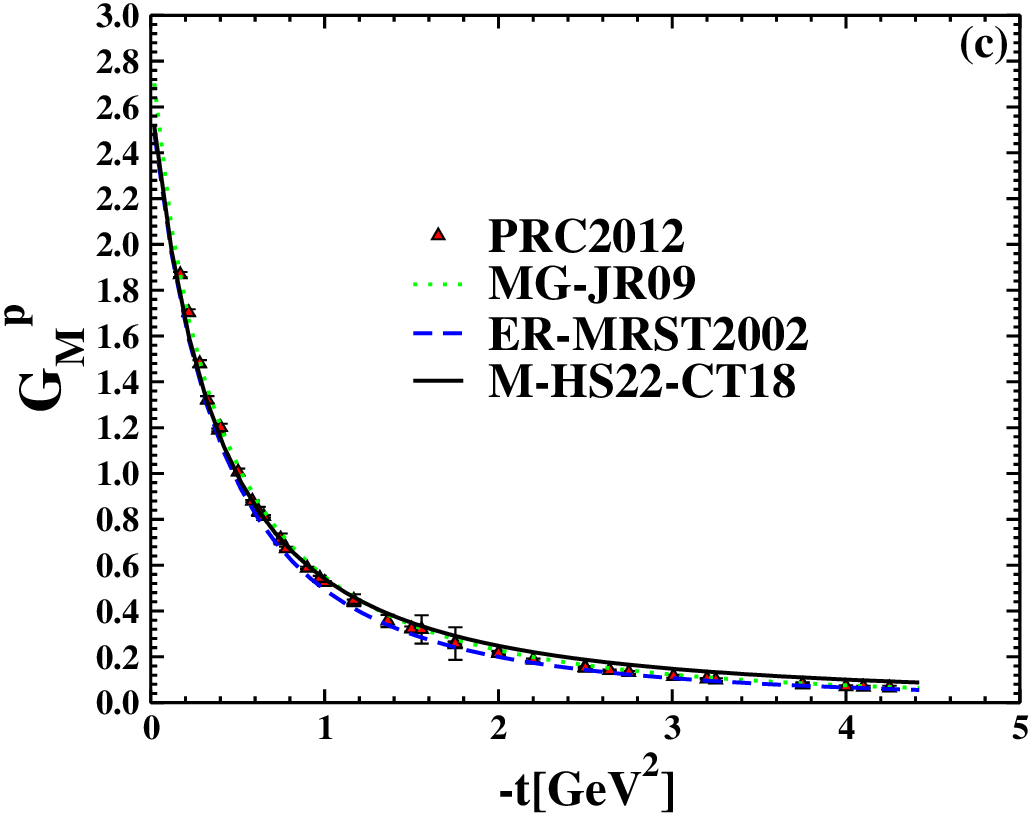}
															\hspace*{2mm}
		\includegraphics[clip,width=0.45\textwidth]{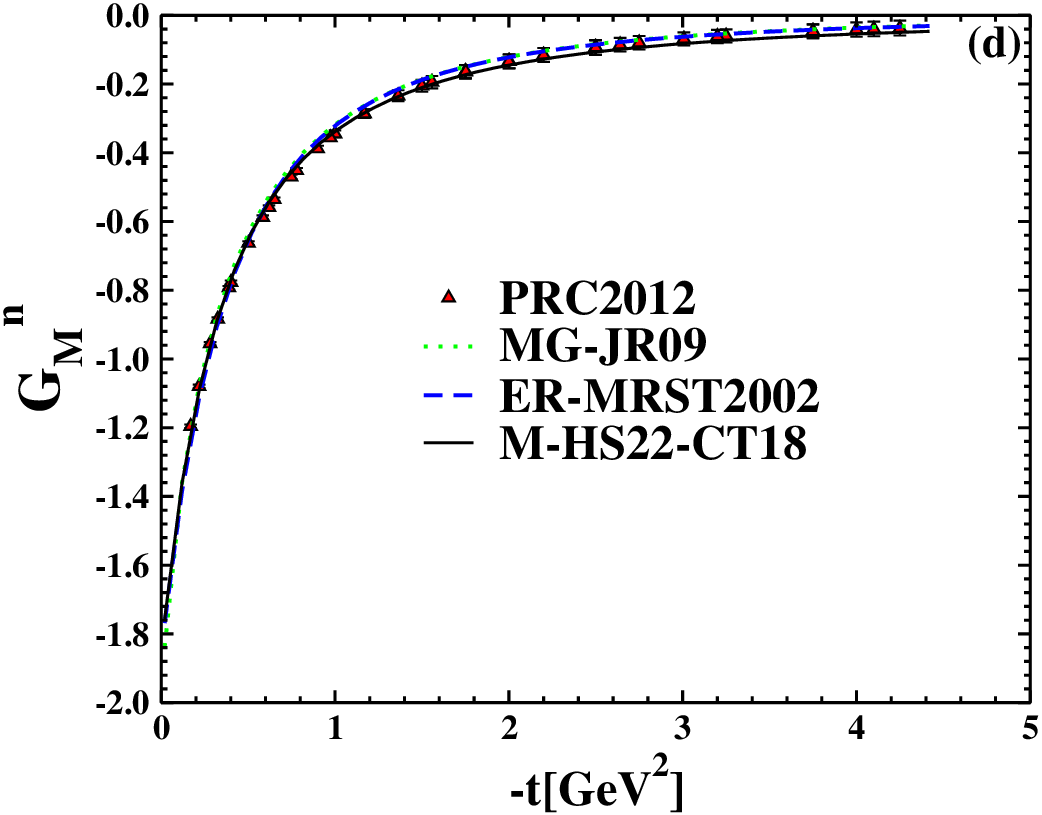}
			\vspace*{1.5mm}
		\caption{\footnotesize    The $G_E^{p,n}$ and $G_M^{p,n}$ as a function of $-t$. Combination of the M-HS22-CT18~\cite{Hou:2019efy}  compare with MG-JR09~\cite{Selyugin:2009ic},~\cite{Jimenez-Delgado:2008orh} and ER-MRST2002~\cite{Guidal:2004nd} ,~\cite{Martin:2002dr}. The points shown are extractions based on experimental data from ~\cite{Qattan:2012zf} (triangle up).}
		\label{fig:GEMpn}
	\end{figure*}

	\begin{eqnarray}
		\varepsilon _{u}(x) &=&\frac{\kappa _{u}}{N_{u}}(1-x)^{\eta _{u}}u_{v}(x), \nonumber\\
		\varepsilon _{d}(x) &=&\frac{\kappa _{d}}{N_{d}}(1-x)^{\eta _{d}}d_{v}(x),\label{eq:Eud1}
	\end{eqnarray}
	\begin{equation}
		\kappa _{q}=\int_{0}^{1}dx\varepsilon _{q}(x).
	\end{equation}
	The limitations on $\kappa_{q}$ are such that the value for the proton must be equal to $F_ 2^P(0)=\kappa_p = 1.793$, and the value for the neutron must be equal to $F_2^n(0)=\kappa_n= -1.913$; 
	\begin{eqnarray}
		\kappa_u=2\kappa_p+\kappa_n,\nonumber\\
		\kappa_d=\kappa_p+2\kappa_n,
	\end{eqnarray}
this gives $\kappa_u = 1.673$ and $\kappa_d = -2.033$. Additionally, the normalization integral for the mathematical constant $\int_{0}^{1}\mathcal{H}_{q}(x,0)$ is equal to $F^P_1(0) = 1$ for the proton, and $F_1^n(0) = 0$ for the neutron. The proton's valence quark number for $u$ and $d$ determines the normalization integrals for the $\mathcal{H}^{u}(x)=u_v(x)$ and $\mathcal{H}^{d}(x)=d_v(x)$ distributions, which are 2 and 1, respectively. The calculated normalization factors $N_u$ and $N_d$ are~\cite{Guidal:2004nd}:
	\begin{eqnarray}
		N_{u} &=&\int_{0}^{1}dx(1-x)^{\eta _{u}}u_{v}(x), \\
		N_{d} &=&\int_{0}^{1}dx(1-x)^{\eta _{d}}d_{v}(x),  \nonumber
		\label{eq:Nud}
	\end{eqnarray}
	
	To satisfy the conditions in Eq.(\ref{eq:Eud1}), the values of $\eta_u$ and $\eta_d$ need to be determined by fitting the nucleon form factor data.
We will now introduce several ansatzes that incorporate $t$ dependence, including the extended ER ansatz~\cite{Guidal:2004nd}, modified Gaussian (MG) ansatz~\cite{Selyugin:2009ic}, and HS22 ansatz~\cite{HajiHosseiniMojeni:2022tzn}, which will be used in this paper.
	
The extended Regge ansatz (ER) is presented by addressing the $t$-dependency of GPDs~\cite{Guidal:2004nd,Nikkhoo:2015jzi}:
	\begin{equation}
		\mathcal{H}^{q}(x,t)=q_{v}(x)x^{-\alpha ^{\prime }(1-x)t},
		\label{eq:HER}
	\end{equation}
	\begin{equation}
		\varepsilon _{q}(x,t)=\varepsilon _{q}(x)x^{-\alpha ^{\prime }(1-x)t},
		\label{eq:EER}
	\end{equation}
	The modified Gaussian ansatz (MG) is also presented, which is~\cite{Selyugin:2009ic,Selyugin:2014sca,Sharma:2016cnf}:
	\begin{equation}
		\mathcal{H}^{q}(x,t)=q_{v}(x)\exp \left[ \alpha \frac{(1-x)^{2}}{x^{m}}t%
		\right] ,
		\label{eq:HMG}
	\end{equation}
	
	\begin{equation}
		\varepsilon _{q}(x,t)=\varepsilon _{q}(x)\exp \left[ \alpha \frac{(1-x)^{2}}{x^{m}}t%
		\right] ,
		\label{eq:EMG}
	\end{equation}
	where the Drell-Yan-West equation is solved by using equations (\ref{eq:HER}) and (\ref{eq:HMG})~\cite{Drell:1969km,West:1970av}. The values of the free parameters for ER and MG are $\alpha^{\prime}=1.15$ and $\alpha=1.09$ and $m=0.45$, respectively.

	And the HS22 ansatz is presented as~\cite{HajiHosseiniMojeni:2022tzn}:
	
	\begin{equation}
		\mathcal{H}_{q}(x,t)=q_{v}\exp [-\alpha^{\prime \prime} t(1-x)\ln (x)+\beta x\ln (1-bt)],
		\label{eq:H}
	\end{equation}
	\begin{equation}
		\varepsilon _{q}(x,t)=\varepsilon _{q}(x)\exp [-\alpha^{\prime \prime} t(1-x)\ln (x)+\beta
		x\ln (1-bt)],
		\label{eq:E}
	\end{equation}
	in which  $\alpha ^{\prime\prime }$, $b$ and $\beta$ are free parameters that are taken respectively as 1.15 , 0.5 and 2.
	
	By changing the parameters in the last ansatz to $\alpha ^{\prime\prime }$= 1.125 , $\beta$= 0.185 and  $b$=1.82 , that have been taken to reproduce the experimental data of  ~\cite{Qattan:2012zf,Cates:2011pz,Diehl:2013xca}, we introduce it as a modified HS22 ansatz (M-HS22) in this paper.

	Nevertheless, we use three distinct parton distributions in our study to calculate the results, \textit{i.e.,}\\

	The MRST2002 global fit at $Q^2_0=1GeV^2$ in the NNLO approximation~ \cite{Martin:2002dr}:
	\begin{eqnarray}
		u_v(x,Q^2_0) &=& 0.262x^{-0.69}(1 - x)^{3.5}\nonumber\\
		&&\times (1 + 3.83\sqrt{x}+ 37.65 x),
		\label{eq:xuvHS22}
	\end{eqnarray}
	\begin{eqnarray}
		d_v(x,Q^2_0) &=& 0.061x^{-0.65}(1 - x)^{4.03}\nonumber\\
		&&\times (1 + 49.05\sqrt{x} + 8.65x).
	\end{eqnarray}

	The JR09 parton distribution functions in the NNLO approximation are mentioned below for a range of input $Q^2_0 = 0.55 GeV^2$~\cite{Jimenez-Delgado:2008orh}:\\
	\begin{eqnarray}
		u_v(x,Q^2_0) &=& 4.4049x^{-0.2125}(1 - x)^{3.6857}\nonumber\\
		&&\times (1 - 1.1483\sqrt{x}+ 4.5921x),
	\end{eqnarray}
	\begin{eqnarray}
		d_v(x,Q^2_0) &=& 13.824x^{-0.1778}(1 - x)^{5.6754}\nonumber\\
		&&\times (1 - 2.2415\sqrt{x} + 3.5917x).
	\end{eqnarray}

	The CT18 parton distribution functions in the NNLO approximation are discussed below for a range of input $Q_0 = 1.3 GeV$~\cite{Hou:2019efy}:   
	\begin{eqnarray}
		{xu_v(x,Q_0^2)}&=&3.385 x^{0.763}(1-x)^{3.036}{pu_v};\nonumber\\
		{pu_v}&=&\text{Sinh}(1.502)(1-y)^4+\nonumber\\
		&&\text{Sinh}(-0.147)4 y(1-y)^3+\nonumber\\
		&&\text{Sinh}(1.671)6 y^2(1-y)^2+\nonumber\\
		&&\left(1+\frac{1}{2}0.763\right) 4 y^3(1-y)+y^4,
		\label{eq:xuvct18}
	\end{eqnarray}
	\begin{eqnarray}
		{xd_v(x,Q_0^2)}&=&0.49 x^{0.763}(1-x)^{3.036}{pd_v};\nonumber\\
		{pd_v}&=&\text{Sinh}(2.615)(1-y)^4+\nonumber\\
		&&\text{Sinh}(1.828)4 y(1-y)^3+\nonumber\\
		&&\text{Sinh}(2.721)6 y^2(1-y)^2+\nonumber\\
		&&\left(1+\frac{1}{2}0.763 \right)4 y^3(1-y)+y^4,
		\label{eq:xdvct18}
	\end{eqnarray}
	In Eqs.(\ref{eq:xuvct18}),(\ref{eq:xdvct18}) set $y=\sqrt{x}$.
	
 Using the M-HS22 ansatz and three parton distribution functions, we obtained the form factors of $u$ and $d$ quarks based on the above formalism and plotted them in Fig.~\ref{fig:tfud1}. The M-HS22 ansatz, particularly in combination with the CT18 PDF, obviously provides a better fit to the experimental data taken from references ~\cite{Diehl:2013xca,Cates:2011pz,Qattan:2012zf}. Additionally, by combining four different types of ansatz with the CT18 PDF and performing calculations, we exhibit the form factors of $u$ and $d$ quarks, the proton and neutron, as well as $t^2 F_1$ and $t^2 F_2$ for the proton and neutron as functions of $-t$ in Figs. \ref{fig:tfud},\ref{fig:fpn}, and ~\ref{fig:t2fud}, respectively.
	
The combination of the M-HS22 ansatz and CT18 PDF for proton and, especially, neutron form factors can provide better agreement with the experimental data than the other combinations mentioned above.

	The Sachs electric and magnetic form factors are expressed  
	from $F_1^N$ and $F_2^N$ combination as~\cite{Qattan:2012zf,Ernst:1960zza}:
	\begin{equation}
		G_{E}^{N}(t)=F_{1}(t)+\frac{t}{4M^{2}}F_{2}(t)    ,\;\;\;G_{M}^{N}(t)=F_{1}(t)+F_{2}(t).
		\label{eq:GN}
	\end{equation}

 The nucleon's charge and magnetic moment are represented by the variables $G_E$ and $G_M$ in the limit $t \rightarrow 0$. Although the neutron electric form factor, $G^n_E$, is almost zero, it has been shown that $G^p_E$, $G^p_M$, and $G^n_M$ roughly follow the dipole form~\cite{Miller:1990iz}.

	 This discovery is in agreement with the simple, nonrelativistic interpretation in which the charge and magnetization of the nucleon are carried by the quarks, and the up and down quarks have similar spatial distributions. For all form factors, the contributions of the up- and down-quark charge distributions are similar, except for $G^n_E$, where there is a nearly complete cancellation between the up- and down-quark charge distributions. The measurements of $G^n_E$ were important in proving that the up-quark distribution differs from the down-quark distribution.
	
	 We can explain how the up and down quarks contribute to the form factors of the nucleon~\cite{Beck:2001yx},
	\begin{eqnarray}
		G^p_{E,M}&=&\frac{2}{3}G^u_{E,M}-\frac{1}{3}G^d_{E,M},\nonumber\\
		G^n_{E,M}&=&\frac{2}{3}G^d_{E,M}-\frac{1}{3}G^u_{E,M}.
	\end{eqnarray}
The expression for the up- and down-quark contributions to the proton form factors is as follows:
	\begin{eqnarray}
		G^u_{E,M}&=&G^n_{E,M}+2G^p_{E,M},\nonumber\\
		G^d_{E,M}&=&2G^n_{E,M}+G^p_{E,M}.
	\end{eqnarray}

	In this convention, $G^u_{E,M}$ reflects the contribution of up quarks in the proton and down quarks in the neutron to the form factors, with analogous formulas for $F_1$ and $F_2$. It is assumed that the quark magnetic moments are the limit of the magnetic form factors as $Q^2$ approaches zero, with $\mu_u = (2\mu_p + \mu_n) =
	3.67\mu_N$ and $\mu_d = (\mu_p + 2\mu_n) = -1.03\mu_N$, respectively. Note that the up- and down-quark contributions, as defined here, include both quark and antiquark contributions and represent the difference between the quark and antiquark distributions due to the charge weighting of the quark and antiquark contributions to the form factors~\cite{Perdrisat:2006hj}.
	
	In Fig.~\ref{fig:GEMpn}, we presented the electric and magnetic form factors of the proton and neutron obtained by employing various combinations of ansatz models and parton distribution functions (PDFs). We have then plotted these form factors as a function of $-t$.

	The following is a calculation of the nucleon Dirac mean squared radii based on the ER~\cite{Guidal:2004nd}: 
	\begin{equation}
		r_{1,p}^{2}=-6\alpha ^{\prime
		}\int_{0}^{1}dx[e_{u}u_{v}(x)+e_{d}d_{v}(x)](1-x)\ln (x),\label{eq:rpREEGE}
	\end{equation}
	\begin{equation}
		r_{1,n}^{2}=-6\alpha ^{\prime
		}\int_{0}^{1}dx[e_{u}d_{v}(x)+e_{d}u_{v}(x)](1-x)\ln (x),\label{eq:rnREEGE}
	\end{equation}
additionally, the electric radii of the proton and neutron are calculated based on the MG model:
	\begin{equation}
		r_{1,p}^{2}=6\alpha \int_{0}^{1}dx[e_{u}u_{v}(x)+e_{d}d_{v}(x)]\frac{%
			(1-x)^{2}}{x^{m}},\label{eq:rPMG}
	\end{equation}
	\begin{equation}
		r_{1,n}^{2}=6\alpha \int_{0}^{1}dx[e_{u}d_{v}(x)+e_{d}u_{v}(x)]\frac{%
			(1-x)^{2}}{x^{m}}.\label{eq:rPMG}
	\end{equation}

In Table(\ref{tab:tabR}), we have calculated nucleon electric radii by using different GPDs and comparing them with experimental data taken from \cite{ParticleDataGroup:2010dbb}.
	\captionsetup{belowskip=0pt,aboveskip=0pt}
	\begin{table}[h]
	\begin{center}
		\caption{{\footnotesize The electric radii of the proton and neutron were calculated using different parton distribution functions (PDFs) based on the extended (ER) \cite{Guidal:2004nd} and  (MG) \cite{Selyugin:2009ic} models. The data used in this study are obtained from \cite{ParticleDataGroup:2010dbb}.}
			\label{tab:tabR}}
			\vspace*{1.5mm}
		\begin{tabular}{ccc}
			\hline\hline\
			\\	PDFs & {\hspace{10mm}}$r_{E,p}$ & {\hspace{10mm}}$r^2_{E,n} $   \\   \hline
			\\Exprimental data   &{\hspace{10mm}}   0.877 $fm$ &{\hspace{5mm}} -0.1161 $fm^2$ \\
			MG-JR09   &{\hspace{10mm}}   0.942 $fm$ &{\hspace{5mm}} -0.1559 $fm^2$ \\   
			MG-MRST2002   &{\hspace{10mm}}   0.900 $fm$ &{\hspace{5mm}} -0.1004 $fm^2$ \\ 
			MG-CT18   &{\hspace{10mm}} 0.880 $fm$ &{\hspace{5mm}} -0.1152 $fm^2$ \\ 
			ER-JR09   &{\hspace{10mm}} 0.857 $fm$ &{\hspace{5mm}} -0.1401 $fm^2$ \\
			ER-MRST2002   &{\hspace{10mm}}   0.839 $fm$ &{\hspace{5mm}} -0.1055 $fm^2$ \\
			ER-CT18   &{\hspace{10mm}}   0.882 $fm$ &{\hspace{5mm}} -0.1159 $fm^2$ \\\\ \hline\hline						
		\end{tabular}
	\end{center}
\end{table}
	\section{TRANSVERSE CHARGE AND MAGNETIZATION
		DENSITIES}\label{sec:sec3}
	The impact parameter distribution of unpolarized quarks inside a nucleon is related to the generalized parton distribution $H$. However, in the case of a nucleon with transverse polarization, the distortion of the quark distribution in the transverse plane is described by $E$~\cite{SattaryNikkhoo:2018gzm}.

	 The distribution of partons in the transverse plane is explained by the two-dimensional Fourier transform of  generalized  parton distribution functions  ~\cite{Guidal:2004nd,Miller:2007uy,Miller:2010nz,Mondal:2015uha}:\newpage
	\begin{eqnarray}
		\rho _{ch}(b) &=&\int \frac{d^{2}q_{\bot }}{(2\pi )^{2}}F_{1}(q^{2})e^{iq_{%
				\bot }\cdot b_{\bot }} \\
		&=&\int_{0}^{\infty }\frac{dQ}{2\pi }QJ_{0}(Qb)F_{1}(Q^{2}),
		\label{eq:ro}
		\nonumber
	\end{eqnarray}
	where $b$ specifies the impact parameter and $J_0$ represents the order zero cylindrical Bessel function. The magnetization density can be obtained from the Fourier transform of the Pauli form factor:
	
	\begin{eqnarray}
		\widetilde{\rho }_{M}(b) &=&\int_{0}^{\infty }\frac{d^{2}q_{\bot }}{%
			(2\pi )^{2}}F_{2}(q^{2})e^{iq_{\bot }\cdot b_{\bot }} \\
		&=&\int_{0}^{\infty }\frac{dQ}{2\pi }QJ_{1}(Qb)F_{2}(Q^{2}),
		\label{eq:rotilda}
		\nonumber
	\end{eqnarray}
	whereas
	\begin{equation}
		\rho _{m}(b)=-b\frac{\partial \widetilde{\rho }_{M}(b)}{\partial b}%
		=b\int_{0}^{\infty }\frac{dQ}{2\pi }Q^{2}J_{1}(Qb)F_{2}(Q^{2}),
		\label{eq:rom}
	\end{equation}
	where the anomalous magnetization density is represented by $\rho _{m}(b)$. Experimental data for transverse densities is not available. However, estimates have been made for the charge and magnetization densities of the proton based on experimental data for electromagnetic form factors \cite{Venkat:2010by,Nikkhoo:2019uqi}.
	
In this section, we have calculated the charge and magnetization densities of protons using the M-HS22 ansatz and three types of parton distribution functions (PDFs), the results are presented in Fig.\ref{fig:roud}, which shows the densities as a function of $b$. We have also used the equations for $G_E$ and $G_M$ from Ref.\cite{Venkat:2010by}, which were obtained from experimental data to compute the transverse charge and magnetization density.

	\section{Impact parameter space and positive constraints for GPDs}\label{sec:sec4}
When the skewness parameter $\xi$ is set to zero, parton distributions in impact parameter space can be related to generalized parton distributions (GPDs) through a simple Fourier transform in transverse momentum  $q_{\bot}$~\cite{Shojaei:2015oia,Burkardt:2003ck}: 
	\begin{figure*}
		\includegraphics[clip,width=0.45\textwidth]{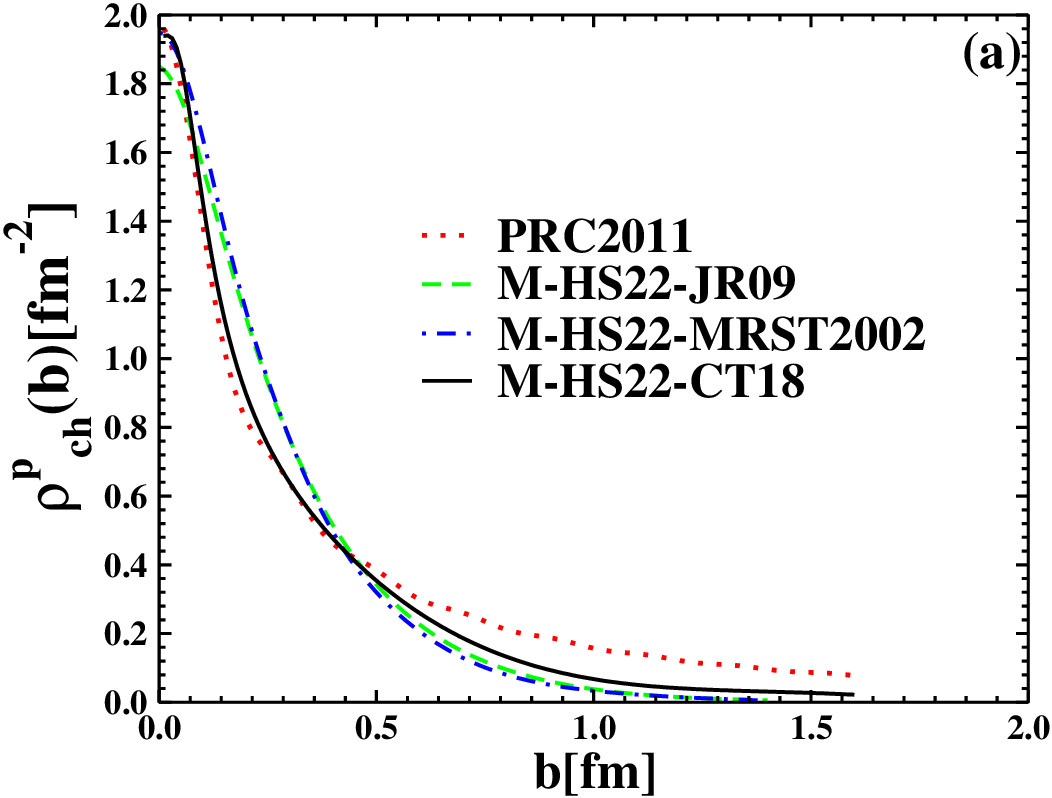}
				\hspace*{2mm}
		\includegraphics[clip,width=0.45\textwidth]{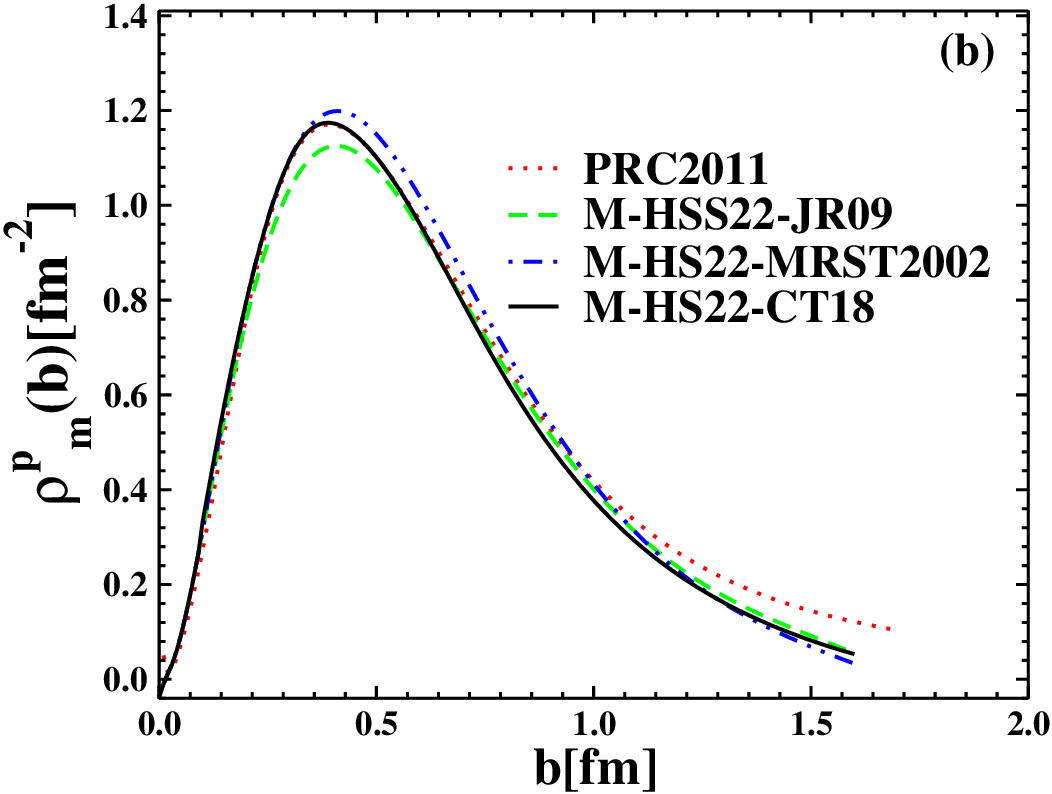}
			\vspace*{1.5mm}
		\caption{\footnotesize   Transverse charge and magnetization densities of the $p$. The M-HS22-CT18~\cite{Hou:2019efy} compare with the M-HS22-JR09~\cite{Jimenez-Delgado:2008orh} and M-HS22-MRST2002~\cite{Martin:2002dr}. The red dotted curve is calculated from the information in~\cite{Venkat:2010by}. }
		\label{fig:roud}
	\end{figure*} 
	\begin{figure*}
		\includegraphics[clip,width=0.45\textwidth]{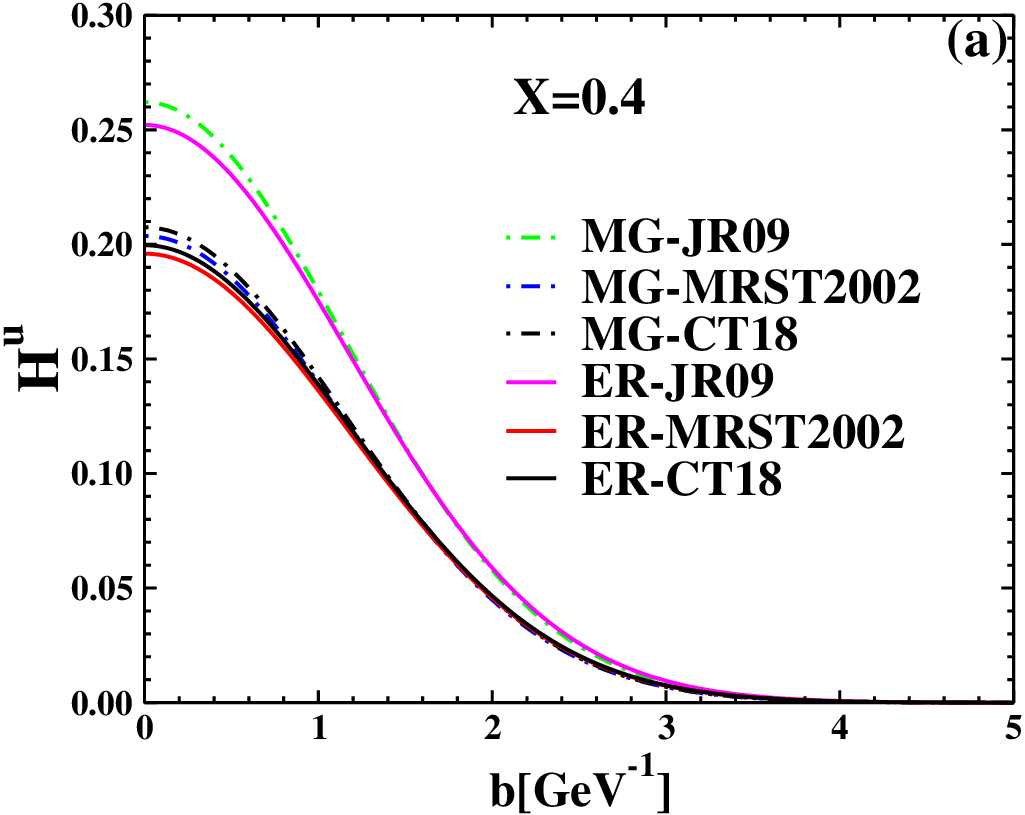}
						\hspace*{2mm}
		\includegraphics[clip,width=0.45\textwidth]{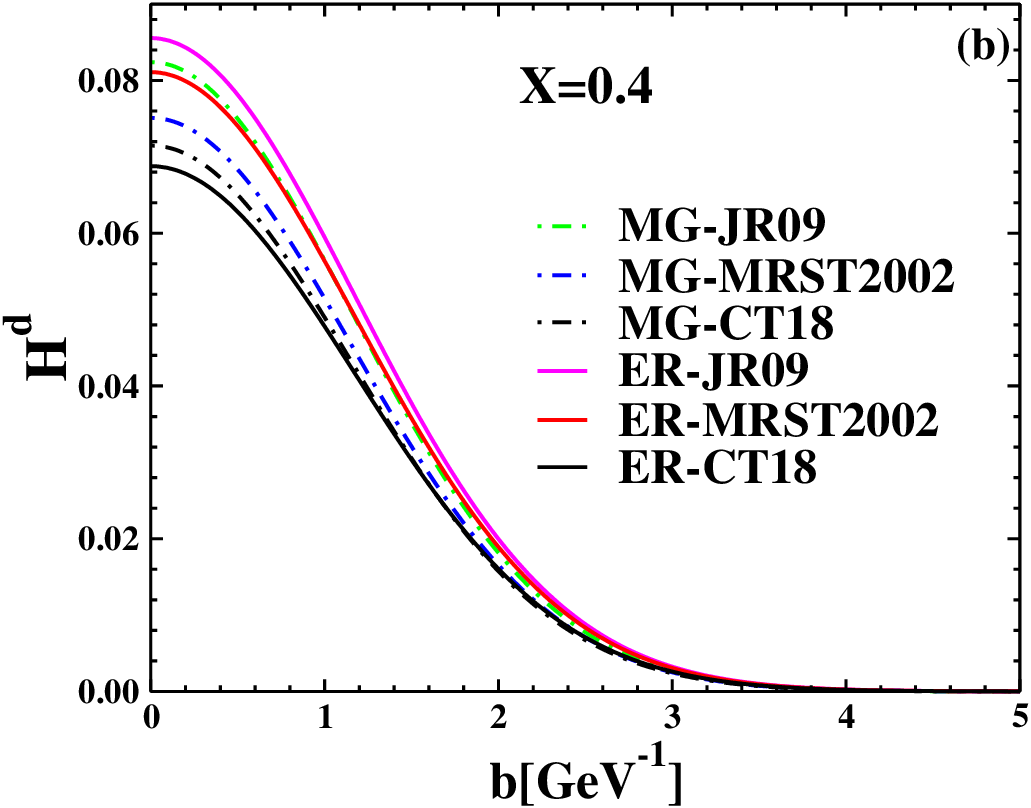}
											\vspace*{3mm}\\
		\includegraphics[clip,width=0.45\textwidth]{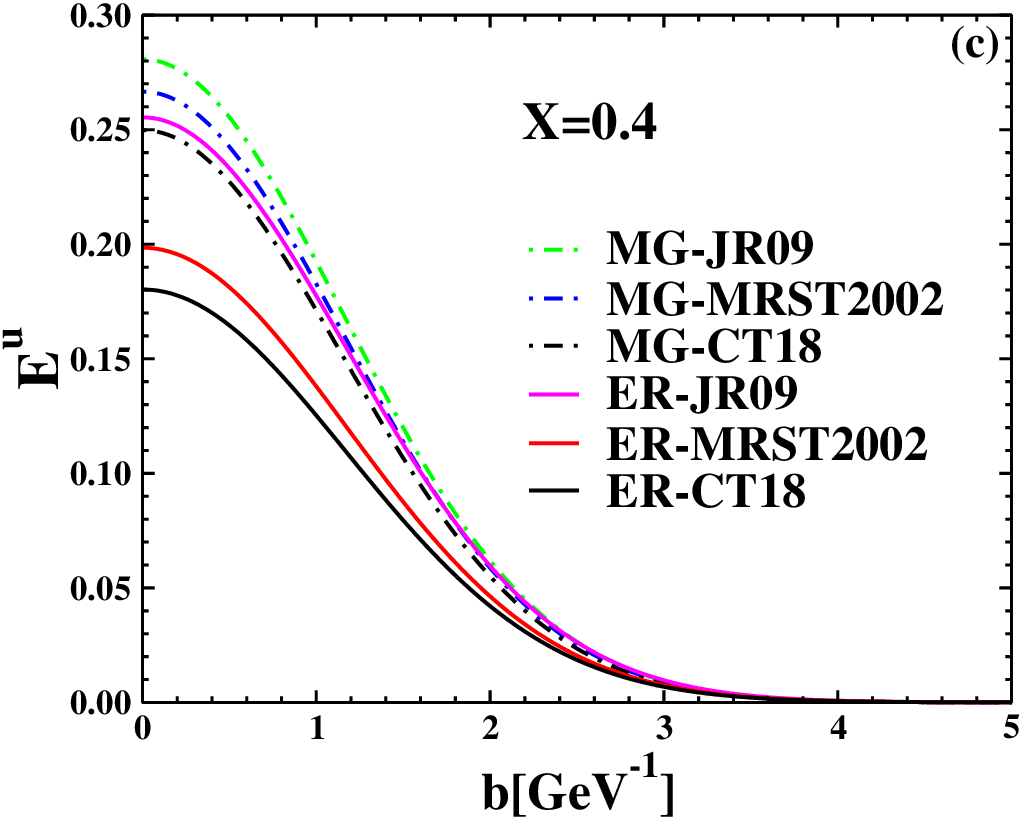}
						\hspace*{2mm}
		\includegraphics[clip,width=0.45\textwidth]{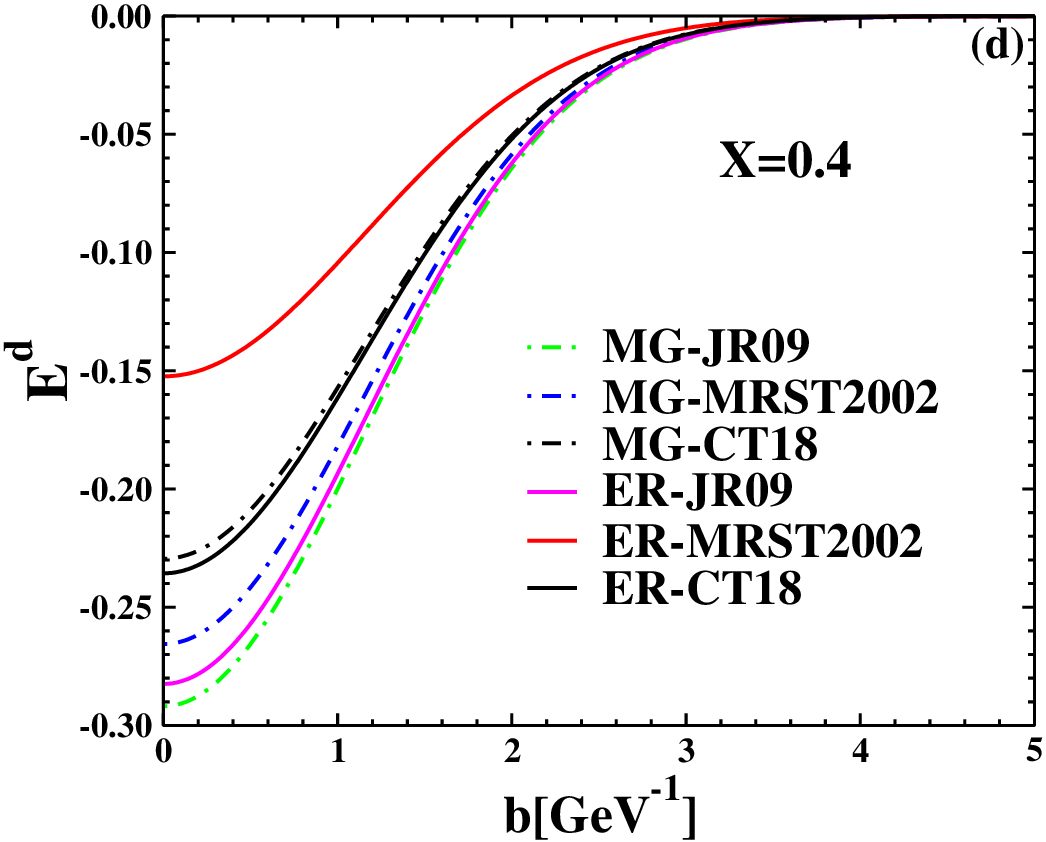}
		\vspace*{1.5mm}
		
		\caption{\footnotesize   $H^q(x,b_{\bot })$ and $E^q(x,b_{\bot })$ for the quarks u and d by combining two different ansatzes, \textit{i.e.} ER~\cite{Guidal:2004nd}  , MG~\cite{Selyugin:2009ic} and PDFs of the JR09~\cite{Jimenez-Delgado:2008orh}, MRST2002~\cite{Martin:2002dr}, and CT18~\cite{Hou:2019efy}.  }
		\label{fig:HEud}
	\end{figure*}
	\begin{equation}
		{H}_{q}(x,b_{\bot })=\int \frac{d^{2}q_{\bot }}{(2\pi )^{2}}%
		e^{ib_{\bot }\cdot q_{\bot }}\mathcal{H}_{q}(x,-q_{\bot }^{2}),
	\end{equation}

	\begin{equation}
		E_{q}(x,b_{\bot })=\int \frac{d^{2}q_{\bot }}{(2\pi )^{2}}e^{ib_{\bot }\cdot
			q_{\bot }}\varepsilon _{q}(x,-q_{\bot }^{2}).
	\end{equation}
	
	This function calculates the probability of detecting a quark in a nucleon with a longitudinal momentum fraction of $x$ at a transverse location of $b_{\bot}$. Based on the positivity constraints, the $E$ and $\mathcal{H}$ GPDs in the impact parameter space can be related as follows~\cite{Miller:2006tv,Miller:1997ya,Miller:1986mk}:
	\begin{equation}
		\frac{1}{2M_{N}}\left\vert \nabla _{b_{\bot }}E_{q}(x,b_{\bot })\right\vert
		\leq {H}_{q}(x,b_{\bot }).\label{eq:eq32}
	\end{equation}

	We can obtain parton distribution functions (PDFs) in the impact parameter space for u and d quarks based on the ER model ~\cite{Guidal:2004nd}:
	\begin{equation}
		{H}_{q}(x,b_{\bot })=q_{v}(x)\frac{e^{-b_{\bot }^{2}/[-4\alpha
				^{\prime }(1-x)\ln (x)]}}{4\pi \lbrack -4\alpha ^{\prime }(1-x)\ln (x)]},
	\end{equation}
	
	\begin{equation}
		E_{q}(x,b_{\bot })=\frac{\kappa _{q}}{N_{q}}(1-x)^{\eta _{q}}q_{v}(x)\frac{%
			e^{-b_{\bot }^{2}/[-4\alpha ^{\prime }(1-x)\ln (x)]}}{4\pi \lbrack -4\alpha
			^{\prime }(1-x)\ln (x)]}.
	\end{equation}
	
	The process of obtaining the PDFs in the impact parameter space for the MG model is as follows:
	
	\begin{equation}
		{H}_{q}(x,b_{\bot })=q_{v}(x)\frac{e^{-b_{\bot }^{2}/[4\alpha
				(1-x)^{2}/x^{m}]}}{4\pi \lbrack 4\alpha (1-x)^{2}/x^{m}]},
	\end{equation}
	
	\begin{equation}
		E_{q}(x,b_{\bot })=\frac{\kappa _{q}}{N_{q}}(1-x)^{\eta _{q}}q_{v}(x)\frac{%
			e^{-b_{\bot }^{2}/[4\alpha (1-x)^{2}/x^{m}]}}{4\pi \lbrack 4\alpha
			(1-x)^{2}/x^{m}]}.
	\end{equation}
	
	Thus, for ER and MG ansatzes, respectively, we have:
	\begin{eqnarray}
		\left\vert \nabla _{b_{\bot }}E_{q}(x,b_{\bot })\right\vert  &=&\frac{\kappa
			_{q}}{N_{q}}(1-x)^{\eta _{q}}q_{v}(x)\frac{b_{\bot }}{2} \\
		&&\times \frac{e^{-b_{\bot }^{2}/[-4\alpha ^{\prime }(1-x)\ln (x)]}}{%
			[-4\alpha ^{\prime }(1-x)\ln (x)]^{2}},  \nonumber
	\end{eqnarray}
	
	\begin{eqnarray}
		\left\vert \nabla _{b_{\bot }}E_{q}(x,b_{\bot })\right\vert  &=&\frac{\kappa
			_{q}}{N_{q}}(1-x)^{\eta _{q}}q_{v}(x)\frac{b_{\bot }}{2}\nonumber \\
		&&\times \frac{e^{-b_{\bot }^{2}/[4\alpha (1-x)^{2}/x^{m}]}}{[4\alpha
			(1-x)^{2}/x^{m}]^{2}}.
	\end{eqnarray}
	Figure~\ref{fig:HEud} depicts the $H$ and $E$ GPDs for $u$ and $d$ quarks, obtained using the JR09 PDF~\cite{Jimenez-Delgado:2008orh}, MRST2002~\cite{Martin:2002dr}, CT18~\cite{Hou:2019efy}, as well as two models: the ER~\cite{Guidal:2004nd} and MG~\cite{Selyugin:2009ic} ansatzes. It is evident from the figure that the ansatzes are more effective in calculations compared to the PDFs.
	
	\section{Conclusion}\label{sec:conclusion}
	In this paper, various ansatzes and PDFs were presented. The Pauli, Dirac, and electromagnetic form factors were calculated using three models of ansatzes MG, ER, and HS22) and three parton distribution functions (MRST2002, JR09, and CT18) for high momentum transfer ranges. The results obtained from all the considered ansatzes and PDFs were systematically compared with each other to parameterize one set of form factors.

	 The free parameters of the HS22 ansatz were modified and introduced as the M-HS22 ansatz. After introducing the formalism, we first selected the M-HS22 ansatz and paired it with different PDFs such as MRST2002, CT18, and JR09. The Dirac and Pauli form factors of the $u$ and $d$ quarks are displayed in Fig.~\ref{fig:tfud1},
	\textit{i.e.}, $F_1^u$, $F_1^d$, $F_2^u$, and $F_2^d$ multiplied by t as a function of $-t$. It can be observed that the M-HS22 ansatz, particularly when used with the CT18 PDF, outperforms other ansatzes and is more consistent with the experimental data presented in~\cite{Cates:2011pz, Diehl:2013xca, Qattan:2012zf}.
 In the next step, we set the CT18 PDF and four ansatzes: HS22, ER, MG, and finally the M-HS22 ansatz, which is shown in Fig.~\ref{fig:tfud}. The proton and neutron's Dirac and Pauli form factors are seen in Fig. \ref{fig:fpn}. It can be seen that the M-HS22 ansatz is more consistent with experimental data than other ansatzes, particularly for the neutron form factor.
 Figure~\ref{fig:t2fud} illustrates the behavior of the $t^2F_1^{p,n}$ and $t^2F_2^{p,n}$ form factors as a function of $-t$. From the last three figures, we can infer that the M-HS22 ansatz exhibits proper behavior and is consistent with experimentally obtained electromagnetic form factors using data from ~\cite{Qattan:2012zf,Cates:2011pz,Diehl:2013xca}. Furthermore, since the combination of CT18 PDF with the MHS22 ansatz is suitable for both protons and neutrons, it yields better results compared to the other combinations. Table~\ref{tab:tabR} shows the computed electric mean squared radius for the nucleons using several ansatzes and PDFs. The result obtained using the CT18 PDF is in agreement with the experimental data reported in\cite{ParticleDataGroup:2010dbb}. Figure~\ref{fig:GEMpn} illustrates the electric form factors of the neutron and proton. It is evident that the parametrizations used for the M-HS22 ansatz agree fairly well with the experimental results. We compared the last two charts with data from \cite{Qattan:2012zf}, and it is easy to conclude from the plotted figures that other combinations of the mentioned ansatzes and PDFs do not have a better correlation with the experimental data than the M-HS22-CT18 combination. In Fig.\ref{fig:roud}, we display the proton charge and magnetization densities for M-HS22 and various PDFs. These densities are computed by considering different parametrizations and comparing them with previous research. To calculate the transverse charge and magnetization density, we employed the $G_E$ and $G_M$ equations based on experimental data from Ref. \cite{Venkat:2010by}. We have presented the GPDs for the ER and MG ansatzes in the impact parameter space in Fig.~\ref{fig:HEud}, taking into account three distinct PDFs. The chosen parametrizations provide a suitable match for the nucleon form factors. This figure is motivated by the fact that a change in ansatz parameters has a greater effect on the results than the influence of PDF. The proposed combination yields a more effective agreement with the Dirac and Pauli form factors of the nucleon compared to other combinations.



	%
	
\end{document}